\documentclass[prd,final,aps]{revtex4}

\usepackage{graphicx}
\usepackage{amsmath,amssymb}

\usepackage{yfonts}
\usepackage{accents}

\newcommand{\lin}{Ref.~\cite{lineareigen}\ }

\newcommand{\g}{\textgoth{g}}
\newcommand{\dP}{\textgoth{P}}
\newcommand{\dL}{\textgoth{L}}

\newcommand{\h}[1][\relax]{\ifx#1\relax \bar{h}  \else  \underaccent{\mathsf{#1}}{\h}  \fi}

\newcommand{\tPsi}[1][\relax]{\ifx#1\relax  {\widetilde{\Psi}}  \else  \underaccent{\,\mathsf{#1}}{\tPsi}  \fi}

\usepackage{amssymb}

\makeatletter

\newcommand{\mathbox}[2][normal]{\begingroup\ensuremath
   \let\@nomath\@gobble  \mathversion{#1}
   \mathchoice
     {\hbox{$\m@th\displaystyle      #2$}}
       {\hbox{$\m@th\textstyle         #2$}}
       {\hbox{$\m@th\scriptstyle       #2$}}
       {\hbox{$\m@th\scriptscriptstyle #2$}}
   \endgroup}

\DeclareFontShape{OT1}{cmss}{bx}{sl}{<-> cmssbxo10}{}

\DeclareMathVersion{boldoblique}
\SetSymbolFont{operators}{boldoblique}{OT1}{cmr}{bx}{sl}
\SetSymbolFont{letters}{boldoblique}{OML}{cmm}{b}{it}
\SetSymbolFont{symbols}{boldoblique}{OMS}{cmsy}{b}{n}
\SetMathAlphabet\mathsf{boldoblique}{OT1}{cmss}{bx}{sl}
\SetMathAlphabet\mathit{boldoblique}{OT1}{cmr}{bx}{it}

\makeatother

\newcommand\vn{{\mathbox[bold]{\mathsf{n}}}}

\begin{document}

\bibliographystyle{prsty} 

\title{The periodic standing-wave approximation: 
post-Minkowski computations}

\author{Christopher Beetle}
\affiliation{Department of Physics,
Florida Atlantic University, Boca Raton, Florida 33431}  

\author{Benjamin Bromley}
\affiliation{Department of Physics,
University of Utah, Salt Lake City, Utah 84112}  
\author{Napole\'on Hern\'andez}
\affiliation{Department of Physics \& Astronomy and Center for 
Gravitational Wave Astronomy, University of Texas at Brownsville,
Brownsville, TX 78520}

\author{Richard H.~Price} 
\affiliation{Department of Physics \& Astronomy and Center for 
Gravitational Wave Astronomy, University of Texas at Brownsville,
Brownsville, TX 78520}

\begin{abstract}
\begin{center}
{\bf Abstract}
\end{center}
The periodic standing wave method studies circular orbits of compact
objects coupled to helically symmetric standing wave gravitational
fields. From this solution an approximation is extracted for the
strong field, slowly inspiralling motion of black holes and binary
stars. Previous work on this model has dealt with nonlinear scalar
models, and with linearized general relativity. Here we present the
results of the method for the post-Minkowski (PM) approximation to
general relativity, the first step beyond linearized gravity.  We
compute the PM approximation in two ways: first, via the standard
approach of computing linearized gravitational fields and constructing
from them quadratic driving sources for second-order fields, and
second, by solving the second-order equations as an ``exact''
nonlinear system.  The results of these computations have two distinct
applications: (i)~The computational infrastructure for the ``exact''
PM solution will be directly applicable to full general relativity.
(ii)~The results will allow us to begin supplying initial data to
collaborators running general relativistic evolution codes.

\end{abstract}
\maketitle

\section{Background and introduction}\label{sec:intro} 
The inspiral of binary black holes is of great interest as a source of
detectible gravitational waves and the computation of the waves from
the inspiral has been the focus of much effort.  Recent breakthroughs
in numerical relativity
\cite{utbpuncture,goddardpuncture} hold the promise of
computing the evolution of the last few orbits of inspiral of a binary
pair of black holes. What remains is to find results for the epoch of
inspiral earlier than the last few orbits, and to provide optimal
initial data for the evolution equations.

The Periodic Standing Wave (PSW) project is intended to fill this gap
in a more-or-less efficient way. This method seeks a numerical
solution for a pair of sources (black holes, neutron stars) in
nondecaying circular orbits with  gravitational fields that are
rigidly rotating, that is, fields that are helically
symmetric. Because the universality of gravitation will not permit
outgoing waves and nondecaying orbits, the solution to be computed is
that for standing waves. An approximation for slowly decaying orbits
with outgoing radiation is then extracted from that numerical
solution.

This work has progressed through several stages. In the first
stage\cite{WKP,WBLandP, rightapprox,paperI,eigenspec}, a nonlinear
scalar fields model was investigated, and numerical methods were
developed to deal with the special mathematical features that would be
common to all standing-wave, helically symmetric computations.  These
features include: (i) a mixed boundary value problem (regions of the
domain in which the equations are hyperbolic and other regions in
which they are elliptic); (ii) an iterative construction of nonlinear
standing wave solutions; (iii) the extraction from the standing wave
solution of an approximate outgoing wave solution; (iv) the
effectiveness of Newton-Raphson methods to deal with the
nonlinearities.  In Ref.~\cite{paperI} standard finite-difference
methods were used to explore the nonlinear scalar problem, but it was
apparent that sufficient resolution to achieve good convergence of the
nonlinear iterations would involve a computationally intensive
project, something we wanted to avoid. Reference \cite{eigenspec}
introduced a new technique for greatly reducing the computational
burden.  That reference introduced ``adapted coordinates'' that were
well suited to the geometry of the problem. Near each of the sources
these coordinates approached spherical coordinates centered on the
source; far from the sources, the coordinates approached standard
spherical coordinates centered on the center of mass. In
Ref.~\cite{eigenspec} it was shown that with these coordinates good
results could be computed by keeping only a very small number of
multipoles, typically just the monopole and quadrupole moments.

With the mathematical and computational methods for scalar fiels under
some control we turned to linearized gravity in the harmonic
gauge\cite{lineareigen}.  The goal in that work was to describe
linearized gravity with convenient ``helical scalars'' (functions only
of coordinates corotating with the helical Killing congruence).  We
presented a formalism that was remarkably simple. Metric perturbations
in the harmonic gauge were described as fields in a Minkowski
spacetime, three complex fields $\widetilde{\Psi}^{(n1)}$,
$\widetilde{\Psi}^{(21)}$, $\widetilde{\Psi}^{(22)}$, and four real
fields $\widetilde{\Psi}^{(nn)}$, $\widetilde{\Psi}^{(n0)}$,
$\widetilde{\Psi}^{(00)}$, $\widetilde{\Psi}^{(20)}$.  With this
description, each of the equations of linearized general relativity,
for each of fields, was found to have the form
\begin{equation}
L \Bigl( \widetilde{\Psi}^{(\mu\nu)} \Bigr)=0
\end{equation}
outside sources.  For the four real fields
$\widetilde{\Psi}^{(\mu\nu)}$, the operator $L$ is simply the
second-order operator of linear scalar theory. For the three complex
fields, $L$ has extra terms.  One of the extra terms turns out to be
imaginary, so that the real and imaginary parts of the complex fields
$\widetilde{\Psi}^{(n1)}$, $\widetilde{\Psi}^{(21)}$,
$\widetilde{\Psi}^{(22)}$ are coupled by the equations. But this is the
only coupling that occurs in this infrastructure for helically symmetric
linearized general relativity.  The presence of the sources is
introduced in Ref.~\cite{lineareigen} through inner boundary
conditions on a small, approximately spherically surface in the source
region.

Here we take the gravitational problem beyond linear theory, by
truncating at second order a nonlinear expansion of the vacuum general
relativistic field equations.  We take two distinctly different views
of the post-Minkowski problem that results. The first, to be called
the ``perturbative post-Minkowski'' (PPM) problem, is the standard approach 
to a sequence of perturbation orders. In this approach, the field equations
are written in the form
\begin{eqnarray}
L \Bigl( \tPsi[1]^{(\mu\nu)} \Bigr)&=&0\label{order1}\\
L \Bigl( \tPsi[2]^{(\mu\nu)} \Bigr)&=&S^{(\mu\nu)}
\Bigl( \tPsi[1]^{(\mu\nu)} , \tPsi[1]^{(\mu\nu)} \Bigr)\label{order2}\,.
\end{eqnarray}
The notation here indicates that one is first to solve the linearized
equations for the first-order perturbative fields
$\tPsi[1]^{(\mu\nu)}$. One then constructs effective
sources for the second-order fields
$\tPsi[2]^{(\mu\nu)}$\,. These effective sources are quadratic in
the linear fields $\tPsi[1]^{(\mu\nu)}$\,. The inner
boundary conditions (representing the physical sources) must only be correct to
linear order in solving Eq.~(\ref{order1}) for
$\tPsi[1]^{(\mu\nu)}$\,.  In Eq.~(\ref{order2}) only the
second-order part of the boundary value must be used for
$\tPsi[2]^{(\mu\nu)}$\,.

The other approach to the second-order post-Minkowskian
solution is to replace Eqs.~(\ref{order1}) and (\ref{order2}), by a
single equation
\begin{equation}\label{exactPM}
L \Bigl( \widetilde{\Psi}^{(ab)} \Bigr)
=S^{(ab)}
\Bigl( \widetilde{\Psi}^{(ab)},\widetilde{\Psi}^{(cd)}
\Bigr)\,.
\end{equation}
In this formulation, the field equations of general relativity are
truncated at second order in the field strengths, and the resulting
system, quadratic in the fields, is treated as a nonlinear field
theory and solved as such. In this ``exact post-Minkowski'' (exact PM) approach,
in contrast to that of Eq.~(\ref{order2}), there is no {\it a priori}
division of $\widetilde{\Psi}^{(ab)}$ into first- and second-order
parts, and the boundary conditions for $\widetilde{\Psi}^{(ab)}$
include both the first- and the second-order parts.

We follow both approaches here. The PPM approach has the
advantage that no nonlinear equations must be solved. All problems of
convergence of a nonlinear solution are therefore avoided. The exact PM
approach has the advantage that it {\em does} require a nonlinear
solution and that it will help to build the computational
infrastructure for full general relativity. 
It will be shown below that, in principle,
the \textit{technical} step 
is surprisingly small
from exact PM field  
equations to those of full general relativity. 
However, there are several important \textit{conceptual} difficulties  
that must be addressed in order to develop a fully general- 
relativistic periodic standing-wave approximation.  These issues will  
be discussed at length in a forthcoming paper \cite{bhgr} by one of us (CB).

The present paper focuses on the technical and computational
challenges of the transition to general relativity, but it cannot
ignore conceptual issues entirely.  This is because two central
conceptual difficulties of that transition, the problems of satisfying
the gauge condition and of computing the source motion, arise already
at second order in our post-Minkowski expansion.  In fact, these two
problems are closely intertwined and, in the second-order problem at
least, can be addressed completely at the analytical level before a
single line of code is written.  We therefore limit our discussion
here to just those aspects of gauge conditions and source motion that
are relevant to the second-order theory.

The field equations (\ref{exactPM}) arise by assuming the harmonic  
gauge condition in the Einstein equation.  This eliminates several  
troublesome terms and yields an equation that, in its linearized  
form at least, can be solved using  standard Green-function  
techniques.  However, a solution of the gauge-fixed field equation  
will only solve the original Einstein equation if it happens to  
satisfy the harmonic gauge condition.  This is not guaranteed.  One  
can show \cite{bhgr} that a necessary condition for a solution of 
Eq.~(\ref{exactPM}) to satisfy the harmonic gauge condition is that the  
binary point sources generating the field should satisfy conservation  
of energy-momentum, $\nabla_a\, T^{ab} = 0$, to second order in  
perturbation theory.  In general relativity, of course, this  
condition naturally incorporates the interaction of each source   
with the other because the derivative operator $\nabla_a$ depends on  
the gravitational field.  As a result, unlike Newtonian gravity,  
conservation of energy in general relativity dictates the dynamics of  
the sources.  That is, $\nabla_a\, T^{ab} = 0$ implies a relativistic  
generalization of Kepler's law, which in Newtonian theory can be  
written in the form
\begin{equation}\label{NewtKep}
\frac {GM}{4ac^2} = \frac{a^2\Omega^2
}{c^2}.
\end{equation}
This result can be considered to be the lowest-order approximation to
a general-relativistically correct formula for some carefully defined
mass as a function of $a$ and $\Omega $.  Note, however, that to
derive even the Newtonian equation of motion, it was necessary to 
calculate fields to
second order in $M$.

To simplify the computations here we make an additional
assumption, though one that is appropriate to the binary
configurations to which the PSW approximation applies: along with our
post-Minkowski approximaton on field strengths, we make a
post-Newtonian-like expansion in orbital velocity. In particular we
consider the $v^2/c^2$ on the right-hand side of Eq.~(\ref{NewtKep}) to
be small, and we keep only the dominant terms in this parameter.
This simplification will apply only to the inner boundary conditions
we use and to some of the details of the equations used for numerical
computation. Aside from these points, the methods developed here are
independent of this low-velocity approximation.

The present paper will make frequent reference to the infrastructure built up
in previous papers, in particular in Refs.~\cite{eigenspec} and \cite{lineareigen}.
The details will not be repeated here, but it will be useful at the outset
to point out the connection to and differences from the Minkowski background
theories of previous papers and the approach in the present paper. It will
also be important to 
define several coordinate systems closely related to those of 
 Refs.~\cite{eigenspec,lineareigen}.
We will assume that 
there exist coordinates $t,x,y,z$ 
that cover the region  of the manifold outside the sources, the 
region in which we shall do computation. 
For convenience, we shall call this system our Minkowski-like 
coordinate system, although 
in the present paper we do not really consider a Minkowski background.
In terms of these 
 $t,x,y,z$
coordinates, we assume that  the Killing
vector of our helical symmetry has the form
\begin{equation}\label{xidef2} 
\xi=\partial_t+\Omega\left(x\partial_y-y\partial_x
\right)
\end{equation}
where $\Omega$
is a constant.

We introduce three
other coordinate systems that are also related to those in
Ref.~\cite{lineareigen}. In that paper they were alternative
coordinates for the Minkowski background. Here they are defined only
as specific transformations of the coordinates $t,x,y,z$. The first of
these is the system $\widetilde{t}=t, \widetilde{x},
\widetilde{y}, \widetilde{z}=z$, defined by
\begin{equation}
\widetilde{x}\equiv x\,\cos{\Omega t}+y\sin{\Omega t}
\quad\quad
\widetilde{y}\equiv \,-x\sin{\Omega t}+y\cos{\Omega t}\,.
\end{equation}
Since ${\cal L}_{\xi}\widetilde{x}={\cal L}_{\xi}\widetilde{y}={\cal
L}_{\xi}\widetilde{z}=0$, the
$\widetilde{x},\widetilde{y},\widetilde{z}$ are labels on trajectories
of the helical Killing congruence and we are justified in calling them
rotating coordinates.  Another set of rotating coordinates is the
cylindrical system $r,z,\varphi $ defined with
$r\equiv\sqrt{x^2+y^2\;}= \sqrt{\widetilde{x}^2+\widetilde{y}^2\;} $,
and with $\tan{\varphi}\equiv\widetilde{y}/\widetilde{x}$.  It should
be noted that the Killing vector of Eq.~(\ref{xidef2}) can be written
as
\begin{equation}\label{xidef3} 
\xi=\partial_t+\Omega\left(\widetilde{x}\partial_{\widetilde{y}}
-\widetilde{y}\partial_{\widetilde{x}}
\right)=\partial_t+\Omega\partial_\varphi
=
\partial_{\widetilde{t}}\,,
\end{equation}
where the last expression is the derivative with $\varphi
$ held constant. For scalar functions the imposition of helical 
symmetry involves the replacement
\begin{equation}\label{helicalreplacement} 
\partial_t\rightarrow -\Omega\partial_\varphi=-\Omega\left(x\partial_y-y\partial_x\right)
=-\Omega\left(\widetilde{x}\partial_{\tilde{y}}-\widetilde{y}\partial_{\tilde{x}}\right)\ .
\end{equation}

It is useful, as in Refs.~\cite{eigenspec} and \cite{lineareigen}, to
define yet another set of rotating Cartesian-like coordinates
$\widetilde{X},\widetilde{Y},\widetilde{Z}$ as a simple renaming
\begin{equation}\label{wtildeXYZdef} 
\widetilde{Z}=\widetilde{x}\quad
\widetilde{X}=\widetilde{y}\quad
\widetilde{Y}=\widetilde{z}\,.
\end{equation}
Our adapted coordinates $\chi,\Theta,\Phi
$ are related to the  $\widetilde{X},\widetilde{Y},\widetilde{Z}$ system 
by the transformation
\begin{eqnarray}
\widetilde{Z}&=&\sqrt{\frac{1}{2}\left[
a^2+\chi^2\cos{2\Theta}+\sqrt{\left(a^4+2a^2\chi^2\cos{2\Theta}+\chi^4\right)\  }
\right]\ }\label{zof}\\
\widetilde{X}&=&\sqrt{\frac{1}{2}
\left[
-a^2-\chi^2\cos{2\Theta}+
\sqrt{\left(a^4+2a^2\chi^2\cos{2\Theta}+\chi^4\right)\  }
\right]\ }\;\cos{\Phi}\label{xof}\\
\widetilde{Y}&=&\sqrt{\frac{1}{2}
\left[
-a^2-\chi^2\cos{2\Theta}+
\sqrt{\left(a^4+2a^2\chi^2\cos{2\Theta}+\chi^4\right)\  }
\right]\ }\;\sin{\Phi}\,,\label{yof}
\end{eqnarray}
and the inverse transformation
\begin{eqnarray}
\chi&\equiv&
\left\{\left[\left(\widetilde{Z}-a\right)^2+\widetilde{X}^2
+\widetilde{Y}^2\right]
\left[\left(\widetilde{Z}+a\right)^2+\widetilde{X}^2
+\widetilde{Y}^2\right]\right\}^{1/4}\label{chiofXYZ}\\
\Theta&\equiv&
\frac{1}{2}\tan^{-1}\left(\frac{2\widetilde{Z}
\sqrt{\widetilde{X}^2+
\widetilde{Y}^2
\;}}{
\widetilde{Z}^2-a^2-\widetilde{X}^2-\widetilde{Y}^2
}\right)\label{ThetofXYZ}\\
\Phi&\equiv&\tan^{-1}{\left(\widetilde{Y}/\widetilde{X}\right)}
\label{PhiofXYZ}\,.
\end{eqnarray}

The remainder of the paper is organized as follows. In
Sec.~\ref{sec:formalism} we introduce our formal expansion of
Einstein's equations and we discuss the truncation of this system to
second order. Here it is demonstrated how in our formalism the change
from second-order post-Minkowski field equations to those of full
general relativity involves only very minor modifications.  We then
discuss the approximation of small orbital velocity that we will use
in the computations. In this section also, we present the derivation
of second-order correct inner boundary conditions.

Section \ref{sec:helicaladap} presents the formalism underlying a
numerical approach. Since we can compute only ``helical scalars,''
unknowns that are functions only of rotating coordinates, we show how
the techniques introduced in \lin can be extended
to the post-Minkowski equations (and to the full Einstein equations).
In this section we also give the detail necessary for casting the
computational problem in terms of the adapted coordinates that proved
to be very efficient in earlier work\cite{eigenspec,lineareigen}.
Section~\ref{sec:numresults} deals with the numerical solution of
Eq.~(\ref{order2}).   Numerical results are given along with a
discussion of the limits of validity of the results and the importance
of nonlinear contributions.  Section \ref{sec:conc} gives a summary of the step
taken in this paper and relates it to what remains to be done.

Throughout the paper we adhere to the conventions of the text by
Misner {\em et al.}\cite{MTW}.  In particular, we use units in which
$c=G=1$.

\section{The form of Einstein's equations}\label{sec:formalism}

\subsection{The full equations}

We follow here the convenient formulation of Landau and Lifschitz 
\cite{landaulifschitz,MTWsec20.3} for the Einstein equations.
This formulation encodes the geometric information in the densitized 
metric
\begin{equation}\label{gDef}
\g^{\alpha\beta}\equiv \sqrt{|{\det g}|}\, g^{\alpha\beta}\,.
\end{equation}
From this we introduce the Landau-Lifschitz quantities
\begin{equation}\label{LLH} 
	\dP^{\alpha\mu\beta\nu} \equiv \g^{\alpha\beta}\, \g^{\mu\nu} 
- \g^{\alpha\nu}\, \g^{\beta\mu} \\
\end{equation}
and 
\begin{multline} \label{LLfld}
	\dL^{\alpha\beta} \equiv \g^{\alpha\beta}{}_{,\mu }\, 
\g^{\mu \nu}{}_{,\nu} - \g^{\alpha\mu }{}_{,\mu }\, \g^{\beta\nu}{}_{,\nu} 
		+ \textstyle{\frac{1}{2}}\, g^{\alpha\beta}\, g_{\mu \nu}\, 
\g^{\mu \lambda}{}_{,\rho}\, \g^{\nu \rho}{}_{,\lambda} 
- g^{\alpha\mu }\, g_{\nu \lambda}\, \g^{\beta\lambda}{}_{,\rho}\, 
\g^{\nu \rho}{}_{,\mu } - g^{\beta\mu }\, g_{\nu \lambda}\, 
\g^{\alpha \lambda}{}_{,\rho}\, \g^{\nu \rho}{}_{,\mu } \\
{}+ g_{\mu \nu}\, g^{\lambda\rho}\, \g^{\alpha\mu }{}_{,\lambda}\, \g^{\beta\nu}{}_{,\rho} 
		+ \textstyle{\frac{1}{8}} \Bigl( 2\, g^{\alpha\mu }\, g^{\beta\nu} 
- g^{\alpha\beta}\, g^{\mu \nu} \Bigr) 
			\Bigl( 2\, g_{\lambda\rho}\, g_{\gamma\sigma} 
- g_{\lambda\sigma}\, g_{\rho\gamma} 
\Bigr) \g^{\lambda\sigma}{}_{,\mu }\, \g^{\rho\gamma}{}_{,\nu}
\end{multline}
(Our notations differ slightly from those of \cite{MTW}, where $\dP^{\alpha\mu\beta\nu}$ is denoted $\textgoth{H}^{\alpha\mu\beta\nu}$ and the Landau--Lifshitz pseudotensor $\dL^{\alpha\beta}$ is denoted $\textgoth{T}^{\alpha\beta}$.  We prefer our notation in the second case because we associate $\textgoth{T}^{\alpha\beta}$ in a related paper \cite{bhgr} with the true, material stress-energy.  We also reserve $\textgoth{H}^{\alpha\mu\beta\nu}$ for other purposes.)  In terms of these quantities, the Einstein equations $G_{\mu\nu}=8\pi T_{\mu\nu}$
are written as
\begin{equation}\label{LLfeq}
	\dP^{\alpha\mu\beta\nu}{}_{,\mu\nu} = \dL^{\alpha\beta} - 16\pi\sqrt{|{\rm det}g|\;}\, T^{\alpha\beta}.
\end{equation}

Putting the definitions (\ref{LLH}),(\ref{LLfld}) into the left side 
of the field equation (\ref{LLfeq}), and rearranging indices slightly, we find  
\begin{equation}\label{LLlhs}
	\dP^{\alpha\mu\beta\nu}{}_{,\mu\nu} = \g^{\mu\nu}\, \g^{\alpha\beta}{}_{,\mu\nu} 
+ \g^{\alpha\beta}\, \g^{\mu\nu}{}_{,\mu\nu} 
		- \g^{\alpha\mu}{}_{,\mu\nu}\, \g^{\beta\nu} - \g^{\alpha\mu}\, 
\g^{\beta\nu}{}_{,\mu\nu} + 2\, \g^{\alpha\beta}{}_{,\mu}\, \g^{\mu\nu}{}_{,\nu} 
		- \g^{\alpha\mu}{}_{,\nu}\, \g^{\beta\nu}{}_{,\mu} 
- \g^{\alpha\mu}{}_{,\mu}\, \g^{\beta\nu}{}_{,\nu}.
\end{equation}
We take our $t,x,y,z
$
``Minkowski'' coordinates to satisfy the harmonic gauge condition that
\begin{equation}\label{hgGauge}
\g^{\alpha\beta}{}_{,\beta} = 0\,.
\end{equation}
in these coordinates.
This choice greatly simplifies the field equations, which become
\begin{multline} \label{rFeq1}
	\g^{\mu\nu}\, \g^{\alpha\beta}{}_{,\mu\nu} =
		\g^{\alpha\mu}{}_{,\nu}\, \g^{\beta\nu}{}_{,\mu} +
		\textstyle{\frac{1}{2}}\, g^{\alpha\beta}\,
		g_{\mu\nu}\, \g^{\mu\lambda}{}_{,\rho}\,
		\g^{\nu\rho}{}_{,\lambda} - g^{\alpha\mu}\,
		g_{\nu\lambda}\, \g^{\beta\lambda}{}_{,\rho}\,
		\g^{\nu\rho}{}_{,\mu} - g^{\beta\mu}\,
		g_{\nu\lambda}\, \g^{\alpha\lambda}{}_{,\rho}\,
		\g^{\nu\rho}{}_{,\mu} \\
		+ g_{\mu\nu}\, g^{\lambda\rho}\,
		\g^{\alpha\mu}{}_{,\lambda}\, \g^{\beta\nu}{}_{,\rho}
		+ \textstyle{\frac{1}{8}} \Bigl( 2\, g^{\alpha\mu}\,
		g^{\beta\nu} - g^{\alpha\beta}\, g^{\mu\nu} \Bigr)
		\Bigl( 2\, g_{\lambda\rho}\, g_{\gamma\sigma} -
		g_{\lambda\sigma}\, g_{\rho\gamma} \Bigr)
		\g^{\lambda\sigma}{}_{,\mu}\,
		\g^{\rho\gamma}{}_{,\nu}.
\end{multline}
To simplify this result further we collect terms on the right hand side as follows
\begin{multline}\label{rFeq2}
	\g^{\mu\nu}\, \g^{\alpha\beta}{}_{,\mu\nu} = \Bigl[
\delta^\alpha_\tau\, \delta^\beta_\lambda\, \delta^\nu_\phi\, \delta^\kappa_\mu
+ \textstyle{\frac{1}{2}}\, g^{\alpha\beta}\, g_{\tau\lambda}\,
\delta^\nu_\phi\, \delta^\kappa_\mu - 2\, \delta^{(\alpha}_\tau\,
g^{\beta)\nu}\, g_{\phi\lambda}\, \delta^\kappa_\mu + \delta^\alpha_\tau\,
\delta^\beta_\lambda\, g_{\phi\mu}\, g^{\kappa\nu} \\
		+ \textstyle{\frac{1}{8}} \Bigl( 2g^{\alpha\kappa}\, g^{\beta\nu} -
			g^{\alpha\beta}\, g^{\kappa\nu} \Bigr) 
                        \Bigl( 2g_{\tau\lambda}\,
			g_{\phi\mu} - g_{\tau\phi}\, g_{\lambda\mu} \Bigr) \Bigr]
			\g^{\tau\phi}{}_{,\kappa}\, \g^{\lambda\mu}{}_{,\nu}\,,
\end{multline}
and we rewrite this as
\begin{multline}\label{rFeq3}
	\g^{\rho\sigma}\, \g^{\alpha\beta}{}_{,\rho\sigma} = \Bigl[
\delta^{(\alpha}_\rho\, \delta^{\beta)}_\sigma -
\textstyle{\frac{1}{2}}\, g^{\alpha\beta}\, g_{\rho\sigma} \Bigr]
\Bigl[ \delta^\rho_\tau\, \delta^\sigma_\lambda\, \delta^\nu_\phi\,
\delta^\kappa_m - 2\, \delta^\rho_\tau\, g^{\sigma\nu}\,
g_{\phi\lambda}\, \delta^\kappa_\mu + \delta^\rho_\tau\,
\delta^\sigma_\lambda\, g_{\phi\mu}\, g^{\kappa y} -
\textstyle{\frac{1}{2}}\, g^{\rho\sigma}\, g_{\tau\lambda}\,
g_{\phi\mu}\, g^{\kappa\nu} \\
		+ \textstyle{\frac{1}{4}}\, g^{\rho\kappa}\,
g^{\sigma\nu} \Bigl( 2g_{\tau\lambda}\, g_{\phi\mu} - g_{\tau\phi}\,
g_{\lambda\mu} \Bigr) \Bigr] \g^{\tau\phi}{}_{,\kappa}\,
\g^{\lambda\mu}{}_{,\nu}.
\end{multline}
It should be noted that the ordinary covariant metric $g_{\alpha\beta}$
and its inverse $g^{\alpha\beta}$ enter (\ref{rFeq3}) only in complementary
pairs.

We define the inverse $\g_{\alpha\beta}$ of our basic field
$\g^{\alpha\beta}$ by
\begin{equation}\label{gInv}
\g^{\alpha\beta}\, \g_{\beta\gamma} = \delta^\alpha{}_\gamma 
\end{equation}
so that 
\begin{equation}
\g_{\alpha\beta} 
= \frac{1}{\sqrt{-\det g}}\, g_{\alpha\beta}\,.
\end{equation}
Using this definition, we can  rewrite the field equation
in the form
\begin{multline}\label{llFeq}
	\g^{\rho\sigma}\, \g^{\alpha\beta}{}_{,\rho\sigma} = 
\Bigl[ \delta^{(\alpha}_\rho\,
\delta^{\beta)}_\sigma - \textstyle{\frac{1}{2}}\, \g^{\alpha\beta}\, \g_{\rho\sigma} \Bigr]
\Bigl[ \delta^\rho_\tau\, \delta^\sigma_\lambda\, \delta^\nu_\phi\, \delta^\kappa_\mu - 2\,
\delta^\rho_\tau\, \g^{\sigma\nu}\, \g_{\phi\lambda}\, \delta^\kappa_\mu + \delta^\rho_\tau\,
\delta^\sigma_\lambda\, \g_{\phi\mu}\, \g^{\kappa\nu} \\
		- \textstyle{\frac{1}{2}}\, \g^{\rho\sigma}\, \g_{\tau\lambda}\,
		\g_{\phi\mu}\, \g^{\kappa\nu} + \textstyle{\frac{1}{4}}\,
		\g^{\rho\kappa}\, \g^{\sigma\nu} \Bigl( 2\g_{\tau\lambda}\, \g_{\phi\mu} -
		\g_{\tau\phi}\, \g_{\lambda\mu} \Bigr) \Bigr]
		\g^{\tau\phi}{}_{,\kappa}\, \g^{\lambda\mu}{}_{,\nu}.
\end{multline}

We next define $\h^{ab} $ by
\begin{equation}\label{hdef} 
\g^{\alpha\beta}
\equiv \sqrt{-\det \eta}\, \bigl(\eta^{\alpha\beta}-\h^{\alpha\beta} \bigr).
\end{equation}
Note that the determinant factor is unity in the 
coordinates $(t, x, y, z)$ in which we take tensor 
components, but technically is needed to define 
the perturbation tensor field $\h^{\alpha\beta}$.
In terms of this new quantity, 
the field equation can be rewritten as
\begin{multline}\label{llheq}
 \Box\h^{\alpha\beta}\equiv \eta^{\rho\sigma}\,
\h^{\alpha\beta}{}_{,\rho\sigma} = -\Bigl[ \delta^{(\alpha}_\rho\,
\delta^{\beta)}_\sigma - \textstyle{\frac{1}{2}}\, \g^{\alpha\beta}\,
\g_{\rho\sigma} \Bigr] \Bigl[ \delta^\rho_\tau\, \delta^\sigma_\lambda\,
\delta^\nu_\phi\, \delta^\kappa_\mu - 2\, \delta^\rho_\tau\, \g^{\sigma\nu}\, \g_{\phi\lambda}\,
\delta^\kappa_\mu + \delta^\rho_\tau\, \delta^\sigma_\lambda\, \g_{\phi\mu}\, \g^{\kappa\nu} \\
		- \textstyle{\frac{1}{2}}\, \g^{\rho\sigma}\, \g_{\tau\lambda}\,
\g_{\phi\mu}\, \g^{\kappa\nu} + \textstyle{\frac{1}{4}}\, \g^{\rho\kappa}\, \g^{\sigma\nu}
\Bigl( 2\g_{\tau\lambda}\, \g_{\phi\mu} - \g_{\tau\phi}\, \g_{\lambda \mu} \Bigr) \Bigr]
\h^{\tau\phi}{}_{,\kappa}\, \h^{\lambda\mu}{}_{,\nu}+\h^{\rho\sigma}\h^{\alpha\beta}{}_{,\rho\sigma}\,.
\end{multline}
It is this form of the Einstein equations  that we shall use through the 
remaining steps of our program. Equation (\ref{llheq}), along with the definition 
(\ref{hdef}), are to be considered as equations for the unknown 
fields $\h^{\alpha\beta}$. 
Note that so far this equation 
is exact; there has been no reference to 
a split into background and perturbations.

\subsection{The post-Minkowski truncation and our expansion}\label{sec:trunc} 

If we consider $\h^{\alpha\beta}$ a perturbation of Minkowski
spacetime, then the left side of Eq.~(\ref{llheq}),
$\eta^{\rho\sigma}\bar{h}^{\alpha\beta}_{\ \ ,\rho\sigma} $, is linear in
this perturbation while the right hand side is of second (and higher)
order.

Keeping only the linear
terms gives us the equations of linearized general relativity 
\begin{equation}\label{lintheory} 
\Box\h[1]^{\alpha\beta}
=0
\end{equation}
in which we have adopted notation like that in Eqs.~(\ref{order1}),(\ref{order2}).
This is equivalent 
to the usual formulation of linearized general relativity since it is easily
shown that, to linear order, $\h^{\alpha\beta}$ is the familiar trace-reversed metric
perturbation 
\begin{equation}
 -\delta g^{\alpha\beta} +\frac{1}{2}\,
\eta^{\alpha\beta}\, \eta_{\rho\sigma}\, \delta g^{\rho\sigma}
=\eta^{\alpha\mu}\eta^{\beta\nu}\left( \delta g_{\mu\nu} - \frac{1}{2}\, \eta_{\mu\nu}\,
\eta^{\rho\sigma}\, \delta g_{\rho\sigma} \right)\,.
\end{equation}
Thus, the $\h^{\alpha\beta}$ defined in Eq.~(\ref{hdef}) agrees to
first order with the well-established notation for linear perturbation
calculations \cite{MTWsec18.1} that was used in
Ref.~\cite{lineareigen}.  Note, however, that its relation to $\delta g_{\mu\nu}$ 
at higher order in perturbation theory is non-linear.

Keeping terms to first
and second order gives our post-Minkowski approximation:
\begin{equation}\label{generalpm} 
\Box\, \h^{\alpha\beta} = S_{\tau\phi\lambda\mu}^{\alpha\beta\kappa\nu}\
\h^{\tau\phi}{}_{,\kappa}\, \h^{\lambda\mu}{}_{,\nu} +
		\h^{\rho\sigma}\, \h^{a\beta}{}_{,\rho\sigma}\,,
\end{equation}
where
\begin{multline}\label{pmtFeq}
S_{\tau\phi\lambda\mu}^{\alpha\beta\kappa\nu}\equiv 
- \Bigl[ \delta^{(\alpha}_\rho\,
	\delta^{\beta)}_\sigma - \frac{1}{2}\, \eta^{\alpha\beta}\,
	\eta_{\rho\sigma} \Bigr] \Bigl[ \delta^\rho_\tau\,
	\delta^\sigma_\lambda\, \delta^\nu_\phi\, \delta^\kappa_\mu - 2\,
	\delta^\rho_\tau\, \eta^{\sigma\nu}\, \eta_{\phi\lambda}\,
	\delta^\kappa_\mu + \delta^\rho_\tau\, \delta^\sigma_\lambda\,
	\eta_{\phi\mu}\, \eta^{\kappa\nu} \\
		- \frac{1}{2}\, \eta^{\rho\sigma}\, \eta_{\tau\lambda}\,
		\eta_{\phi\mu}\, \eta^{\kappa\nu} + \frac{1}{4}\,
		\eta^{\rho\kappa}\, \eta^{\sigma\nu} \Bigl( 2\,
		\eta_{\tau\lambda}\, \eta_{\phi\mu} - \eta_{\tau\phi}\,
		\eta_{\lambda \mu} \Bigr) \Bigr]\,.
\end{multline}
It should be noted that the conversion of these post-Minkowski
equations to the equations of full general relativity requires only
the replacement of the $\eta$\,s by $\g$\,s in
$S_{\tau\phi\lambda\mu}^{\alpha\beta\kappa\nu} $.  No changes need to
be made in the wave operator on the left or the differentiated fields on the right.
 This means that a computer code designed to
solve Eq.~(\ref{pmtFeq}) can very simply be converted to one that
solves the full theory.

As pointed out in Sec.~\ref{sec:intro}, there are two ways in which
the equations of Eq.~(\ref{generalpm}) and (\ref{pmtFeq}) can be
approached. 
Here we describe the PPM method, the simpler
standard path of solving Eq.~(\ref{lintheory}) first for the $\h$\,s
correct to first order, then using these first-order correct $\h$\,s to
construct a known right hand side of Eq.~(\ref{generalpm}).  In the
notation of Eq.~(\ref{order2}), our problem becomes
\begin{equation}\label{generalpm2} 
\Box\, \h[2]^{\alpha\beta}
= S_{\tau\phi\lambda\mu}^{\alpha\beta\kappa\nu}\
\h[1]^{\tau\phi}{}_{,\kappa}\, \h[1]^{\lambda\mu}{}_{,\nu} +
\h[1]^{\rho\sigma}\, \h[1]^{a\beta}{}_{,\rho\sigma}\,,
\end{equation}
where $\h[1]^{\rho\sigma}$ is a solution of Eq.~(\ref{lintheory}).

To proceed, we must more carefully consider just what the nature is of
our approximation scheme.  In the usual post-Minkowski
theory\cite{vanmeter,crjohnson} an expansion in field strength is
used, and the particle velocities are considered to be of zeroth
perturbative order.  Here we are using a different scheme which is
best thought of as as an expansion in the source mass $M$. A more
careful description of our approximation is that we are considering a
family of helically symmetric solutions of the Einstein equations
describing two co-orbiting ``particles,'' each of mass $M$, moving
opposite to one another on a common circular orbit of radius $a$, and
coupled to standing waves. The parameter $\epsilon=M/a$ is the
parameter on which we base a small parameter expansion. Via Kepler's
law, or a relativistic extension of it, the velocity $v$ of our source
objects is of order $\sqrt{M/a\;}$. It is convenient, therefore, to
consider a factor of $v $ to represent a half-order in our expansion
scheme. A quantity proportional to $vM/a$, for example, will be
considered to be order 1.5.

In this expansion scheme $\h[1]^{\alpha\beta}$ is, in principle, of
first-order in $M/a $ (or second order in $v$). Not all ``linear''
components are of this order, however. The linearized field equations
of general relativity, in the harmonic gauge, are
\begin{equation}\label{linwsource} 
\Box\h[1]^{\alpha\beta}
=-16\pi T^{\alpha\beta}\,,
\end{equation}
and the stress-energy component $T^{tt}$ is proportional to the source
mass $M $ and, to lowest order, is independent of $v$. All other
components of the stress energy are, to lowest order, proportional to
one or more factors of $v $.  Components $T^{ti}$ (where $i$ is a
spatial index) are of order 1.5 in $M$, and components $T^{ij} $ are
of order 2. Thus $\h[1]^{tt}$, in linearized general relativity, is
the only component of $\h[1]^{\alpha\beta}$ that is driven by a
first-order source, and hence is the only component that is actually
of first order.  Here, we summarize the somewhat complicated situation
regarding orders in linearized theory and PPM equations.

(i) For $\bar{h}^{tt}$, the first order fields are found from
linearized gravity, Eq.~(\ref{lintheory}) or (\ref{linwsource}) and
are of order $v^0 M/a$ (that is, first
order). Equation~(\ref{generalpm2}) is used to solve for
$\h[2]^{tt}$, with contributions on the right only from
$\h[1]^{tt}$.

(ii) For $\bar{h}^{ti}$ the lowest order fields are of order 1.5.  To
solve for these fields, we need only use linearized theory, i.e.\,,
Eq.~(\ref{lintheory}) or (\ref{linwsource}). In principle, we could
adapt Eq.~(\ref{generalpm2}), with $\h[2.5]^{ti}$ on the left,
putting on the right products of $\h[1]^{tt}$ and
$\h[1.5]^{ti}$, both from linearized theory. We do not solve for these 
corrections to $\bar{h}^{ti}$ 
in the present paper since
they are higher than second order.  The computation of the lowest
order fields has already been described in Ref.~\cite{lineareigen}.
There Eq.~(\ref{linwsource}) was solved for all components of
$\bar{h}^{\alpha\beta}$, and it was pointed out that the procedure is 
inconsistent, i.e.\,, that the solutions are of different order.

(iii) For $\bar{h}^{ij}$ the lowest order fields are of order 2.
Thus, it is inconsistent to solve Eq.~(\ref{lintheory}) or
(\ref{linwsource}) for these lowest order fields. The consistent
procedure is to use an adaption of Eq.~(\ref{generalpm2}) with
$\h[2]^{ij}$ on the left, and with products of
$\h[1]^{tt}$ on the right. In the case of
Eq.~(\ref{linwsource}), the spatial components of the stress energy
are to be included.  

We should note that our approximation scheme has some of the 
spirit of a post-Newtonian rather than a  post-Minkowskian perturbation
methods. But in our scheme there are waves at every level of
approximation, and there is no $c\rightarrow\infty $ Newtonian
limit. It is, therefore, justifiable to consider our approach to 
be a type of post-Minkowski expansion.

\subsection{The gauge issue}\label{gaugeissues}

To specify a solution, we must, of course, add source terms or
boundary data to Eqs.~(\ref{lintheory}) or (\ref{pmtFeq}). The
solutions thereby determined must satisfy the gauge condition in
Eq.~(\ref{hgGauge}) or equivalently must, in principle, satisfy
$\h^{\alpha\beta}_{\ ,\beta}=0 $.  In practice, this condition is relaxed in a
post-Minkowski approximation. If we are computing the fields correct
to order $n$, then the gauge condition need be satisfied only to order
$n$, that is:
\begin{equation}\label{hGauge} 
\h[n]^{\alpha\beta}{}_{,\beta}={\cal O}\left(\;\h[n+1]^{\mu\nu} \right)\,.
\end{equation}
Thus in linear theory we must only have that $\h[1]^{\alpha\beta}{}_{,\beta} $ 
be of second order, and in the post-Minkowski problem of
Eq.~(\ref{pmtFeq}) we must only have the gauge condition satisfied to 
second order.

For sources moving in binary orbits, this issue is  apparent in
linearized theory and the way in which we deal with it was  raised in
Ref.~\cite{lineareigen}; we review that argument here.  
In our present notation, in the linearized theory, the field equation becomes
that of Eq.~(\ref{linwsource}).
The stress energy tensor for a pair of binary point masses (the stress energy 
used in Ref.~\cite{lineareigen}) will
only satisfy $T^{\alpha\beta}_{\ ,\beta}=0$ if the masses are at rest.
For masses moving at velocity $v
$ there are components of 
$T^{\alpha\beta}_{\ ,\beta}$ that are of order $vM/a$ and of order $v^2M/a$, and the solution to 
Eq.~(\ref{linwsource}), therefore,  cannot satisfy $\h^{ab}_{\ ,b}
=0.$
The way in which we are to understand this is by considering the missing 
terms in 
\begin{equation}\label{linextended} 
\Box\h^{\alpha\beta}
=-16\pi T^{\alpha\beta}+{\cal O}\left(\h^2\right)\,.
\end{equation}
The second-order terms on the right can be thought of as representing
the gravitational forces that drive the source masses. The gauge condition
to lowest nontrivial order  will involve the divergence of the stress 
energy and the quadratic post-Minkowski terms. Satisfying that gauge condition is 
what gives us the relationship of $v
$ and $M/a
$, i.e.\,, it is what gives us Kepler's law. To the order at which we are working, 
keeping terms only of order $(\bar{h}^{\alpha\beta})^2$, we will infer only the 
standard Newtonian Kepler's law. A higher order treatment would lead to relativistic
corrections of the relationship of 
$v
$ and $M/a
$.

\subsection{Boundary conditions}

At distances $r$ much larger than the orbital radius (and therefore
much larger than the source masses $M$) the post-Minkowskian
corrections to the metric are tiny. The boundary conditions on the
fields, then, are just those of linear theory, as described in
\lin. These conditions are the usual Sommerfeld conditions, that in
the Minkowski coordinates all fields $f(t,x,y,z) $ far from the
sources obey $\partial_tf=\partial_rf$ for ingoing waves, and
$\partial_tf=-\partial_rf$ for outgoing waves. (Here $r$ is the radial
coordinate $\sqrt{x^2+y^2+z^2}$.)  In general relativity, of course,
neither ingoing nor outgoing helically symmetric waves are possible,
but these conditions are needed in constructing the standing wave
solutions that are possible in general relativity.

In order to complete the specification of the fields, we must give
the conditions at inner boundaries, i.e.\,, at small coordinate distances $R $
from the location of the sources. Here ``small distances'' means that
$R/a\ll1$ where $R$ is some characteristic distance from the location
of the source.  Within the scope of our approximation we are then
looking for conditions near the source that correspond both to
$R/a\ll1$ and to second order in $M/a $.

The equations for which we seek boundary conditions are those of
Eqs.~(\ref{generalpm}) and (\ref{pmtFeq}).  The right-hand side of
Eq.~(\ref{generalpm}) is known from the solution of the linearized
problem.  The only ambiguity then is what homogeneous solution of
$\Box\bar{h}^{\alpha\beta}=0$ to choose. To lowest order in $M/a$, the
answer is clear: we choose the same moving monopole solution as in
\lin.  The choice of the homogeneous solution to second order in $M/a$
is more subtle. In principle we could have a solution that corresponds
to a moving dipole. For example, we could have
\begin{equation}
\bar{h}^{tt}=\vec{p}\cdot\vec{R}/R^3\,,
\end{equation}
where $\vec{R} $ is the distance, in a local (approximate) Lorentz
frame from a source location to a field point.  The dipole constant
$\vec{p} $ would have to represent the acceleration of the local
Lorentz frame since, for a spherically symmetric source, there is no
other spatial vector related to the physics of the problem.
Dimensionally this second-order term would have to be proportional to
$M^2a^0 $. This would not make physical sense since this
``acceleration term'' would be independent of the orbital radius
$a$. The arguments against second-order homogeneous solutions of
higher multipolarity are an extension of this idea.
It is worth emphasizing that the choice of the inner boundary condition
is equivalent to the choice of the ``particles'' whose motion produces 
the fields. We are free to choose both first- and second-order 
inner boundary conditions to model sources with large intrinsic multipoles. 
Such intrinsic multipoles would be independent of the orbital radius $a$.

The particular solution of Eq.~(\ref{generalpm}) can be found, of course,
from the known form of the right-hand side of the equation. From dimensional
arguments alone, one would expect that in the limit of small distance $R$ from
the source, the second-order solution has the character $M^2/R^2$, 
$M^2/Ra$, and so forth.  We shall impose our inner boundary 
at $R\ll a$, so that the second-order term 
$M^2/Ra$ is negligibly small compared to  $M^2/R^2$.  (For a post-Minkowski
approximation to be valid in the computational region, 
we must, of course, also require that $M/R
$ be small.) We repeat here the conditions on the choice of the parameters 
of the physical configuration and the inner boundary: 
\begin{equation}
  {M}/{R}\ll1\quad\quad\quad{R}/{a}\ll1\,.
\end{equation}
With these choices the inner boundary conditions will be independent
of $a$, and hence the second-order correct boundary conditions can be
found from the Schwarzschild solution expressed in coordinates
$T,X,Y,Z $\cite{notrelated} that satisfy the harmonic gauge
condition\cite{weinberg,vanmeter}
\begin{equation}\label{harmschw}
ds^2=-\left(\frac{1-M/\bar{R}}{1+M/\bar{R}}\right)dT^2
+\left(1+\frac{M}{\bar{R}}\right)^2
\left(dX^2+dY^2+dZ^2
\right)
+\left(\frac{1+M/\bar{R}}{1-M/\bar{R}}\right)
\frac{M^2}{\bar{R}^4}
\left(Xd{Z}+Yd{Y}+{Z}d{Z}
\right)\,,
\end{equation}
where $\bar{R}^2=X^2+Y^2+Z^2$.  (Note: These particle-centered
coordinates are not related to the
$\widetilde{X},\widetilde{Y},\widetilde{Z}$ coordinates of
Eqs.~(\ref{wtildeXYZdef})--(\ref{PhiofXYZ}).)  The next step is to
expand this solution in $M/R$, adopting the method of
Johnson\cite{crjohnson} and Van Meter\cite{vanmeter}.  
We let $z^\mu(\tau)$ be the coordinate path of one of the source
particles, where $\tau $ is the particle proper time, and we let
$U^\mu=dx^\mu/d\tau $ be the 4-velocity of the particle.


We next consider the metric in Eq.~(\ref{harmschw}) to be written as
$\sqrt{|{\det g}|}\, g^{\alpha\beta}
=\eta^{\alpha\beta}-\h^{\alpha\beta}$, as in Eqs.~(\ref{gDef}) and
(\ref{hdef}), so that $\bar{h}^{\alpha\beta}$ is of first and higher
order in $M/\bar{R} $.  For an event at $x^\mu$, we define
\begin{equation}\label{defnullr} 
r^\mu=x^\mu-z^\mu(\tau)
\end{equation}
so that $r^\mu$ is a null vector, and $\tau $ is the retarded time for the event, to
zeroth order in $M/{\bar{R}}$.
Next we define 
\begin{equation}
{\cal R}\equiv -{U}_\alpha r^\alpha
\end{equation}
to be the distance from the particle  to the event, to zeroth order in 
 $M/{\bar{R}}$,
measured in the particle comoving frame,
and we define 
\begin{equation}\label{ndef}
n^\mu=\frac{r^\mu}
{\cal R}
-U^\mu
\end{equation}
to be the spatial direction to the
event at $x^\mu$, to zeroth order, in the particle comoving frame.
In terms of these quantities, the fields near the particles,
correct to second order in $M/{\cal R}
$ are
\begin{equation}\label{hbargen} 
\bar{h}^{\mu\nu}=\left(\frac{4M}{\cal R}+\frac{7M^2}{{\cal R}^2}\right)U^\mu U^\nu
-\frac{M^2n^\mu n^\nu
}{{\cal R}^2}\,.
\end{equation}

As in \lin we choose the sources to be locally-spherical 
points on circular paths\cite{noissue}.
One of the source points (``particle 1'') moves on the path
\begin{equation}
x=a\cos{\Omega t}\quad\quad
y= a\sin{\Omega t}\quad\quad
z=0.
\end{equation}
and the other (``particle 2'') has opposite signs in the formulas for
$x(t)$ and $y(t)$.
The instantaneous 4-velocity  is
\begin{equation}
U^0=\gamma\ \ \ \ \quad\quad U^y=\pm v\gamma\ .
\end{equation}
Here and below the upper sign refers to particle 1, the lower to particle 2.
The particle motion clearly satisfies the helical symmetry, a necessary condition
for the ``particles'' to be sources of helically symmetric fields.

We define $v$ to be the coordinate speed $a\Omega 
$ of the particles, and $\gamma$
to be the associated Lorentz factor $\gamma\equiv1/\sqrt{1-v^2
\;}$. 
To proceed we follow the derivation of Eqs.~(22)-(28) of \lin to relate
the retarded quantities to our Minkowski-like coordinates. This is 
done by introducing instantaneosly corotating coordinates $\widetilde{x},
\widetilde{y},\widetilde{z}
$:
\begin{equation}
y=\widetilde{y}\pm vt\quad\quad
\widetilde{x}=x\quad\quad\widetilde{z}=z\quad\quad\
\widetilde{r}^2\equiv(\widetilde{x}-a)^2+\widetilde{y}^2+\widetilde{z}^2\,.
\end{equation}
For an event
$t,x,y,z $ near (i.e.\,, at small ${\cal R} $ from) a particle it was
shown in \lin  that $t_{\rm part}
$, the retarded time at the particle, is given by
\begin{equation}
t_{\rm part}=t\mp v\gamma^2\tilde{y}-\gamma
\sqrt{\tilde{r}^2+\gamma^2v^2\tilde{y}^2\;}\,.
\end{equation}
From this we have 
\begin{equation}
{\cal R}\equiv\sqrt{\tilde{r}^2+\gamma^2v^2\tilde{y}^2\;}\,,
\end{equation}
and for the null vector of Eq.~(\ref{defnullr})
we have
\begin{equation}
{r}^\mu
=\{t-t_{\rm part}, x\mp a, y\mp vt_{\rm part},z\,.
\end{equation}
With $t_{\rm part}$ eliminated, this is
\begin{equation}\label{vecr} 
{r}^\mu
=\{\pm v\gamma^2\tilde{y}+\gamma{\cal R},\
x\mp a,\ \gamma^2\tilde{y}\pm v\gamma{\cal R},\ z
\}
\end{equation}
and the ${n}^\mu
$ vector of Eq.~(\ref{ndef}) is
\begin{equation}
{n}^\mu
=\{\pm\frac{v\gamma^2\tilde{y}
}{\cal R},\ \frac{x\mp a}{\cal R},\ \frac{\gamma^2\tilde{y}
}{\cal R}, \frac{z}{\cal R}\ \}\ . 
\end{equation}

With these expressions we can now explicitly evaluate the components of 
the inner boundary condition in
Eq.~(\ref{hbargen}):
\begin{eqnarray}
\bar{h}^{tt}&=&\left(\frac{4M}{{\cal R}}\gamma^2+\frac{7M^2}{{\cal R}^2}
\right)\gamma^2
 -\frac{M^2}{{\cal R}^2}
v^2\gamma^4\frac{\tilde{y}^2}{{\cal R}^2}\label{innerBChnn}\\
\bar{h}^{tx}=\bar{h}^{xt}&=&\mp\frac{M^2}{{\cal R}^2}v\gamma^2
\frac{\widetilde{y}(x\mp a)}{{\cal R}^2}\\
\bar{h}^{ty}=\bar{h}^{yt}&=&\pm\left(\frac{4M}{{\cal R}}\gamma^2+\frac{7M^2}{{\cal R}^2}
\right)\,v\gamma^2
 \mp\frac{M^2}{{\cal R}^2}
v\gamma^4
\frac{\tilde{y}^2}{{\cal R}^2}\\
\bar{h}^{tz}=\bar{h}^{zt}&=&\mp\frac{M^2}{{\cal R}^2}
v\gamma^2\frac{z\tilde{y}}{{\cal R}^2}\\
\bar{h}^{xx}&=&-
\frac{M^2}{{\cal R}^2}\,\frac{(x\mp a)^2
}{{\cal R}^2}\\
\bar{h}^{xy}=\bar{h}^{yx}&=&-\frac{M^2}{{\cal R}^2}\,
\gamma^2\;\frac{\tilde{y}(x\mp a)}{{\cal R}^2}\\
\bar{h}^{xz}=\bar{h}^{zx}&=&-\frac{M^2}{{\cal R}^2}\,
\frac{z(x\mp a)}{{\cal R}^2}\\
\bar{h}^{yy}&=&\left(\frac{4M}{{\cal R}}+\frac{7M^2}{{\cal R}^2}
\right)\,v^2\gamma^2
 -\frac{M^2}{{\cal R}^2}
\gamma^4\frac{\tilde{y}^2}{{\cal R}^2}\\
\bar{h}^{yz}=\bar{h}^{zy}&=&-\frac{M^2}{{\cal R}^2}\,
\gamma^2\,
\frac{\tilde{y}z}{{\cal R}^2}\\
\bar{h}^{zz}&=&-\frac{M^2}{{\cal R}^2}\;\frac{z^2}{{\cal R}^2}\,.\label{innerBChzz}
\end{eqnarray}

\section{Equations for computation: helical scalars and 
adapted coordinates}\label{sec:helicaladap}

\subsection{The equations in helical scalar form}\label{sec:helicalscalar}

In \lin the metric perturbation fields were considered to live on a
Minkowski background.  We introduced a set of basis vectors
$\bf{n}=\partial_t $, $\bf{e}_x=\partial_x$, $\bf{e}_y=\partial_y$,
$\bf{e}_z=\partial_z$ that were covariantly constant in that Minkowski
background. In the present paper, and in full general relativity, it
is no longer convenient to consider fields on a Minkowski
background. It is, however, important to use the infrastructure of
\lin for constructing helical scalars.  To do this we will use the
same symbols as in \lin, but with a somewhat different meaning. Here
these quantities are to be interpreted as indexed objects whose
components are constant.Thus, for example $\bf{n}=\partial_t $, is to
have indices $n^t=1$, $n_t=-1$ and $n^j=n_j=0$. The index manipulation
of these symbols will done as if they were components in a Minkowski
basis.

In \lin, we constructed 10, rank 2, symmetric basis tensors ${\bf t}_A$
that were covariantly constant in the Minkowski background.  Here it
is convenient to use the same notation
\begin{eqnarray}
{\bf t}_{nn}&\equiv&{\bf n}{\bf n}\label{tnndef}  \\
{\bf t}_{n0}&\equiv& \textstyle{\frac{1}{\sqrt{2\;}}}\left[{\bf n}{\bf e}_z
+{\bf e}_z{\bf n}
\right]\\
{\bf t}_{n,\pm1}&\equiv&
\textstyle{\frac{\mp1}{{2\;}}}\left[
{\bf n}({\bf e}_x\pm i{\bf e}_y)
+({\bf e}_x\pm i{\bf e}_y){\bf n}
\right]\label{tn1def}
\\
{\bf t}_{0,0}&\equiv&\textstyle{\frac{1}{\sqrt{3\;}}}
\left[{\bf e}_x{\bf e}_x+{\bf e}_y{\bf e}_y+{\bf e}_z{\bf e}_z\right]\\
{\bf t}_{2,0}&\equiv&\textstyle{\frac{-1}{\sqrt{6\;}}}
\left[{\bf e}_x{\bf e}_x+{\bf e}_y{\bf e}_y-2{\bf e}_z{\bf e}_z\right]\\
{\bf t}_{2,\pm1}&\equiv&\mp\textstyle{\frac{1}{2}}
\left[{\bf e}_x{\bf e}_z+{\bf e}_z{\bf e}_x\right]
-\textstyle{\frac{1}{2}}\,i
\left[{\bf e}_y{\bf e}_z+{\bf e}_z{\bf e}_y\right]\\
{\bf t}_{2,\pm2}&\equiv&\textstyle{\frac{1}{2}}
\left[{\bf e}_x{\bf e}_x-{\bf e}_y{\bf e}_y\pm i\left({\bf e}_y{\bf e}_x+{\bf e}_x{\bf e}_y
\right)
\right]\,.\label{t22def}
\end{eqnarray}
Here we intepret these simply to be constant indexed objects with the
numerical values they would have in a Minkowski basis. Thus, for
example, $t_{nn}^{tt}=1$.  Lastly, we define
\begin{eqnarray}
\widetilde{\bf t}_{nn}\equiv{\bf t}_{nn}\quad
\widetilde{\bf t}_{n0}\equiv{\bf t}_{n0}&\quad&
\widetilde{\bf t}_{00}\equiv{\bf t}_{00}\quad
\widetilde{\bf t}_{20}\equiv{\bf t}_{20}\label{ttildefirst}\\
\widetilde{\bf t}_{n,\pm1}\equiv
e^{\mp i\Omega t}
{\bf t}_{n,\pm1}
&\quad&
\widetilde{\bf t}_{2,\pm1}\equiv
e^{\mp i\Omega t}
{\bf t}_{2,\pm1}\\
\widetilde{\bf t}_{2,\pm2}&\equiv&
e^{\mp2i\Omega t}\label{ttildelast}
{\bf t}_{2,\pm2}\,.
\end{eqnarray}
We can express $\bar{h}^{\alpha\beta} $ as a sum using either type of
tensor-like basis:
\begin{equation}\label{tilderepbarh} 
\bar{h}^{\alpha\beta}
=\Psi^{A}{\bf t}^{\alpha\beta}_{\ A}
=\widetilde\Psi^{A}\widetilde{\bf t}^{\alpha\beta}_{\ A}
=\widetilde\Psi^{(00)}\widetilde{\bf t}^{\alpha\beta}_{00}+\ldots{}
+\widetilde\Psi^{(2,-2)}\widetilde{\bf t}^{\alpha\beta}_{2,-2}\,,
\end{equation}
where the label $A$ takes any of the 10 values $(nn)\cdots(2,-2)$.

The components  $\widetilde{t}^{\alpha\beta} _A$ are not constant, but
they have a very useful property. They behave under ${\cal L}_\xi$
like the components of a Lie-dragged tensor. This means that the
scalar-like quantities $\widetilde{\Psi}^A$ constructed in
Eq.~(\ref{tilderepbarh}) will be helical scalars, i.e.\,, they will
be constant along the helical trajectories.
As in Ref.~\cite{lineareigen}, these basis tensors have
the following important property for differentiation with respect to
the Minkowski coordinates
\begin{equation}\label{dBas}
	\partial_\mu\,\widetilde{\bf t}^{\alpha\beta}_A = \partial_\mu
\left( e^{-i\mu(A)\Omega {\bf t}}\, {\bf t}^{\alpha\beta}_A \right) =
i\mu(A)\Omega\, e^{-i\mu(A)\Omega t}\, {\bf t}^{\alpha\beta}_A\, \vn_\mu
= i\mu(A)\Omega\, \widetilde{\bf t}^{\alpha\beta}_A\, \vn_\mu\,,
\end{equation}
where $\mu(A)$ has the value 0 for $A=(nn), (n0), (00), (20) $, has
 the value $\pm1 $ for $A=(n,\pm1), (2,\pm1)$ and has the value $\pm2
 $ for $A=(2,\pm2) $.

The $\widetilde{\Psi}^{A}
$ representation for $\bar{h}^{\alpha\beta}$ can be substituted into
Eq.~(\ref{generalpm}), and the result can be contracted with the (orthogonal)
basis symbols $\widetilde{\bf t}^{\alpha\beta}_{\ A}$. The result
is a set of equations of the form
\begin{equation}\label{boxComplex} 
\Box\widetilde{\Psi}^{A}
-2i\mu(A)\Omega^2\partial_{\varphi}\widetilde{\Psi}^{A}
+\mu(A)^2
\Omega^2\widetilde{\Psi}^{A}=
{\cal Q}^{A}\,.
\end{equation}
For 
$A=(n,\pm1),(2,\pm1),(2,\pm2)$, the fields $\widetilde{\Psi}^A$
are complex, and in practice we work with the real and imaginary 
parts 
\begin{equation}
\widetilde\Psi^{A}=U^{A}+iV^{A}\,,
\end{equation}
of these fields. Separated into its real and imaginary parts,
Eq.~(\ref{boxComplex}) becomes
\begin{eqnarray}
\Box{U}^{A}
+2\mu(A)\Omega^2\partial_{\varphi}{V}^{A}
+\mu(A)^2\Omega^2{U}^{A}&=&\mbox{Real part of }\left({\cal Q}^A\right)\label{BoxU}  \\
\Box{V}^{A}
-2\mu(A)\Omega^2\partial_{\varphi}{U}^{A}
+\mu(A)^2\Omega^2{V}^{A)}&=&\mbox{Imaginary part of }\left({\cal Q}^A\right)\,.\label{BoxV} 
\end{eqnarray}
The $\Box$ operator here, as in Eq.~(\ref{boxComplex}), is
$\eta^{\alpha\beta}\partial_\alpha\partial_\beta$, and the helically
symmetric time derivatives are implemented through the replacement
$\partial_t\rightarrow-\Omega(x\partial_y-y\partial_x)=-\Omega\partial_\varphi
$, so that
\begin{equation}
\Box=\partial_{\tilde{x}}^2+\partial_{\tilde{y}}^2+\partial_{\tilde{z}}^2
-\Omega^2\partial_{\varphi}^2\,.
\end{equation}

In \lin  the right-hand sides ${\cal Q}^A$ were zero in the region
outside stress-energy sources.  In the post-Minkowski approximation of
the present paper, the right hands side follow from the contraction
with $\widetilde{\bf t}^{\alpha\beta}_{\ A}$ of the right-hand side of
the equations in (\ref{generalpm}), (\ref{pmtFeq}), expressed in terms of the
representation of $\bar{h}^{\alpha\beta}$ given in
Eq.~(\ref{tilderepbarh}).  The effective source term ${\cal Q}^A$, then
will consist of terms quadratic in derivatives of the
$\widetilde{\Psi}^{A}$, and [from the last term on the right in
Eq.~(\ref{generalpm})] in products of $\widetilde{\Psi}^{A}$ and its
second derivatives. The form of the source terms then is
\begin{equation}\label{RAdef} 
{\cal Q}^A={\cal S}^{\widetilde{a}\tilde{b} A}_{\ \ \ BC}
\;\partial_{\tilde{a}}\widetilde{\Psi}^B
\;\partial_{\tilde{b}}\widetilde{\Psi}^C
+
{\cal T}^{\widetilde{a}\tilde{b} A}_{\ \ \
BC}\;\widetilde{\Psi}^B
\;\partial_{\tilde{a}}\partial_{\tilde{b}}\widetilde{\Psi}^C
\end{equation}
where the coefficients ${\cal S}^{\widetilde{a}\tilde{b} A}_{\ \ \ BC}$,
${\cal T}^{\widetilde{a}\tilde{b} A}_{\ \ \ BC}$, relatively simple 
functions of the rotating coordinates, are derived in  the appendix.
Equations 
(\ref{boxComplex}) and
(\ref{RAdef})
are the field equations we solve for our post-Minkowski approximation.  

The inner boundary conditions, 
are those of 
Eqs.~(\ref{innerBChnn})--
(\ref{innerBChzz}), converted to projections on the helically symmetric 
basis vectors. A straightforward computation gives
\begin{equation}\label{nonadap-nn} 
\widetilde{\Psi}^{\bf nn}=\left(\frac{4M}{\cal R}+\frac{7M^2}{{\cal R}^2}\right)\gamma^2
\,-\,\frac{M^2
}{{\cal R}^2} \frac{v^2\gamma^4\tilde{y}^2
}{{\cal R}^2
}
\end{equation}
\begin{equation}\label{nonadap-n0} 
\widetilde{\Psi}^{\bf n0}
=\pm\,\frac{M^2}{{\cal R}^2} 
\,\frac{v\gamma^2\tilde{y}z}{{\cal R}^2}
\end{equation}
\begin{equation}\label{nonadap-n1} 
e^{-i\Omega t}\widetilde{\Psi}^{\bf n1}=\mp\,\left(\frac{4M}{\cal R}+\frac{7M^2}{{\cal R}^2}\right)
iv\gamma^2 
\mp\,\frac{M^2
}{{\cal R}^2}\,\frac{v\gamma^2\tilde{y}}{{\cal R}}
\frac{(x\mp a-i\gamma^2
\tilde{y})}{\cal R}
\end{equation}
\begin{equation}\label{nonadap-00} 
\widetilde{\Psi}^{\bf 00}=\left(\frac{4M}{\cal R}+\frac{7M^2}{{\cal R}^2}\right)
 \frac{v^2\gamma^2}{\sqrt 3\;}
-\frac{M^2
}{{\cal R}^2}\,\left(\frac{(x\mp a)^2+\gamma^4\tilde{y}^2+z^2
}{\sqrt{3\;}{\cal R}^2}\right)
\end{equation}
\begin{equation}\label{nonadap-20} 
\widetilde{\Psi}^{\bf 20}=-\,\left(\frac{4M}{\cal R}+\frac{7M^2}{{\cal R}^2}\right)
\,\frac{v^2\gamma^2}{\sqrt{6\;}}
+\frac{M^2}{{\cal R}^2}\,
\frac{(x\mp a)^2+\gamma^4\tilde{y}^2-2z^2}
{{\sqrt{6\;}}{\cal R}^2}
\end{equation}
\begin{equation}\label{nonadap-21} 
e^{-i\Omega t}
\widetilde{\Psi}^{\bf 21}=
+\frac{M^2}{{\cal R}^2}\,
\,\frac{z}{{\cal R}}\frac{(x\mp a-i\gamma^2\tilde{y})}{{\cal R}}
\end{equation}
\begin{equation}\label{nonadap-22} 
e^{-2i\Omega t}\widetilde{\Psi}^{\bf 22}=
-\left(\frac{4M}{\cal R}+\frac{7M^2}{{\cal R}^2}\right)\frac{v^2\gamma^2}{2}
-\,\frac{M^2}{{\cal R}^2}
\left(
\frac{(x\mp a)^2
-\gamma^4\tilde{y}
^2}{2{\cal R}^2}
-i\,\frac{(x\mp a)\gamma^2
\tilde{y}}{{\cal R}^2}
\right)\,.
\end{equation}
Here and below we omit equations for $A=(n,-1)$,
$(2,-1), $ and $(2,-2)$,
since the quantities carrying these labels are (up to a sign) the
complex conjugates of the the quantities with $A=(n1)$,
$(21)$, and $(22)$.

In Sec.~\ref{sec:intro} it was explained that 
our expansion is really based on a family of solutions of Einstein's equations 
with varying $\epsilon=M/a$, with $M
$ the mass of a source, and $a$ the radius of the orbits. We noted that the orbital
velocity, of order $\sqrt{M/a\;}$, can be used to keep track of orders of terms, and 
we noted that of the components  $\bar{h}^{\alpha\beta}
$ only $\bar{h}^{tt}$ is truly first order. 

That argument was based on considerations of stress-energy sources in
linearized gravity.  In the present paper, we represent the effect of the
source objects through inner boundary conditions, rather than through
explicit stress energy sources. In
Eqs.~(\ref{innerBChnn})-(\ref{innerBChzz}) we can see explicitly the
orders of those inner boundary conditions. Again, of the components of
$\bar{h}^{\alpha\beta}$, only $\bar{h}^{tt}$ has a first-order piece
to its inner boundary conditions, a piece that goes as
$(M/a)v^0$. (Note: The factor $M/{\cal R}$ should be viewed as the
order parameter $M/a$, divided by the dimensionless distance ${\cal
R}/a $ that is the limit appropriate to ``near-source'' inner boundary
conditions.) The inner boundary condition for $\bar{h}^{tt}$ also has
a second-order piece $M^2/a^2 $ and a piece $(M^2/a^2)v^2$ that is
third order in $M/a $, or sixth order in $v$. By similar counting, we
see that the lowest order term for $\bar{h}^{ty}$ is $\sim v^3$, while
$\bar{h}^{tx}$ and $\bar{h}^{tz}$ are $\sim v^5$, with all other
components $~v^4$.  For the description of the field in terms of
helical scalars, Eqs.~(\ref{adap-nn}) -- (\ref{adap-22}) tell us that
$\widetilde{\Psi}^{(nn)}$ is of first (and higher) order in $M/a$ or
of order $v^2 $, while $\widetilde{\Psi}^{(n1)}$ is of order $v^3$ and
all other helical scalars are at least of order $v^4$. We summarize here
the leading orders of all $\Psi^{A}$:
\begin{equation}\label{orders} 
\Psi^{(nn)}\sim v^2\quad\quad  \Psi^{(n1)}\sim v^3\quad \quad 
\Psi^{(00)},\Psi^{(20)},\Psi^{(21)},
\Psi^{(22)}\sim v^4\quad\quad  \Psi^{(n0)}\sim v^5\ .
\end{equation}
A consequence of this is that for our second-order post-Minkowski
solutions only the $\widetilde{\Psi}^{(nn)}$ terms need to be kept on
the right in Eq.~(\ref{RAdef}), so that those source-like terms can be
simplified to the order $v^4
$ expression 
\begin{equation}\label{RAsimple} 
{\cal Q}^A={\cal S}^{\widetilde{a}\tilde{b} A}
\;\partial_{\tilde{a}}\widetilde{\Psi}^{(nn)}
\;\partial_{\tilde{b}}\widetilde{\Psi}^{(nn)}
+
{\cal T}^{\widetilde{a}\tilde{b} A}\;\widetilde{\Psi}^{(nn)}
\;\partial_{\tilde{a}}\partial_{\tilde{b}}\widetilde{\Psi}^{(nn)}\,.
\end{equation}
The details of these terms are given in the appendix. What is of greatest
importance is the order of these driving terms:
\begin{equation}\label{drivingorders} 
  Q^{(nn)}\sim v^4 \quad\quad 
  Q^{(n0)},Q^{(n1)}  \sim v^5 \quad\quad 
Q^{(00)},Q^{(20)},Q^{(21)},  Q^{(22)} \sim v^4\,.
\end{equation}
It follows that the computation of the second-order
$\widetilde{\Psi}^{(nn)}$ requires a solution of a nonlinear
problem. To lowest order ($v^3 $) computation of
$\widetilde{\Psi}^{(n1)}$ is linear, and does not require a driving
term. For the other fields ( $\widetilde{\Psi}^{(n0)}$,
$\widetilde{\Psi}^{(00)}$, $\widetilde{\Psi}^{(20)}$,
$\widetilde{\Psi}^{(20)}$,$\widetilde{\Psi}^{(22)}$), the driving term
is of the same order as the lowest order field. In this case, the
solution for the lowest order field requires inclusion of
the driving term. The resulting problem is not nonlinear, however, since
the driving term involves not the field being computed, but rather the
first-order field $\widetilde{\Psi}^{(nn)}$.

\subsection{Adapted coordinates}

The corotating coordinates $\widetilde{x},\widetilde{y},\widetilde{z}$ 
are not well suited to describing the fields and sources of the 
rotating binary. As in Refs.~\cite{eigenspec,lineareigen} we introduce
a set of ``adapted coordinates,'' $\chi,\Theta,\Phi$  transformations
of $\widetilde{x},\widetilde{y},\widetilde{z}$ that are better suited 
to encoding information about the sources and fields, especially in combination
with the truncation of a multipole analysis, as laid out in 
Refs.~\cite{eigenspec,lineareigen}.

The introduction of adapted coordinates imposes two changes in the
details of Eqs.~(\ref{boxComplex}), (\ref{RAdef}). First the
operators $\Box
$ and $\partial_\varphi
$ take the form
\begin{multline}\label{waveq}
\Box\Psi=A_{\chi\chi}\;\frac{\partial^2\Psi}{\partial\chi^2}
+A_{\Theta\Theta}\;\frac{\partial^2\Psi}{\partial\Theta^2}
+A_{\Phi\Phi}\;\frac{\partial^2\Psi}{\partial\Phi^2}
+2A_{\chi\Theta}\;\frac{\partial^2\Psi}{\partial\chi\partial\Theta}
+2A_{\chi\Phi}\;\frac{\partial^2\Psi}{\partial\chi\partial\Phi}
+2A_{\Theta\Phi}\;\frac{\partial^2\Psi}{\partial\Theta\partial\Phi} \\
+B_{\chi}\;\frac{\partial\Psi}{\partial\chi}
+B_{\Theta}\;\frac{\partial\Psi}{\partial\Theta}
+B_{\Phi}\;\frac{\partial\Psi}{\partial\Phi}\,,
\end{multline}
\begin{equation}\label{FDMoutbc} 
\frac{\partial}{\partial\varphi}=
\left(\Gamma^\Theta\frac{\partial}{\partial\Theta}
+\Gamma^\Phi\frac{\partial}{\partial\Phi}
+\Gamma^\chi\frac{\partial}{\partial\chi} \right)\,,
\end{equation}
where the coefficients $A,B,\Gamma$ are known functions of $\chi,\Theta,\Phi$
given explicitly in the appendix of Ref.~\cite{lineareigen}.
The second change needed is that the derivatives in the expressions
for the ${\cal Q}^A$ must be converted to derivatives with respect to
the adapted coordinates. Since the rotating coordinates are relatively
simple functions of the adapted coordinates, this change is
straightforward; the details are given in the appendix.

Though each step of the transformation to adapted coordinates involves
elementary functions, the full set of steps that must be taken in
transforming, and in imposing the eigenspectral method and multipole
filtering\cite{eigenspec} becomes exceedingly tedious and prone to
error.  For this reason, the transformations, eigenspectral method,
and the generation of the final finite difference equations for computation have been
implemented as symbolic manipulation with {\em Maple}.

With the helical symmetry 
of Eq.~(\ref{helicalreplacement}), 
and with the fact that far from the sources $\chi\rightarrow r$, 
the Sommerfeld outer boundary conditions, $\partial_t\widetilde{\Psi}^A
=\pm\partial_\varphi
\widetilde{\Psi}^A$,
becomes
\begin{equation}\label{sommerfeldhelical}
\partial_\chi\widetilde{\Psi}^A=\pm\Omega 
\left(\Gamma^\Theta\frac{\partial\widetilde{\Psi}^A}{\partial\Theta}
+\Gamma^\Phi\frac{\partial\widetilde{\Psi}^A}{\partial\Phi}
+\Gamma^\chi\frac{\partial\widetilde{\Psi}^A}{\partial\chi} \right)\,,
\end{equation}
where + and - correspond respectively to outgoing and ingoing conditions.

The inner boundary conditions of Eqs.~(\ref{nonadap-nn})-(\ref{nonadap-22})
become
\begin{equation}\label{adap-nn} 
\widetilde{\Psi}^{\bf nn}=\left(\frac{4M}{\cal R}+\frac{7M^2}{{\cal
R}^2}\right)\gamma^2
\,-\,\frac{M^2
}{{\cal R}^2} \frac{v^2\gamma^4
}{{\cal R}^2}
\frac{\chi^4}{4a^2}
\sin^2{(2\Theta)}\cos^2{\Phi}
\end{equation}
\begin{equation}\label{adap-n0} 
\widetilde{\Psi}^{\bf n0}
=\pm\,\frac{M^2}{{\cal R}^2} 
\,\frac{v\gamma^2}{{\cal R}^2}
\frac{\chi^4}{4a^2}
\sin^2{(2\Theta)}\sin{\Phi}\cos{\Phi}
\end{equation}
\begin{multline} \label{adap-n1} 
\widetilde{\Psi}^{\bf n1}=\left[-\,\left(
\frac{4M}{\cal R}+\frac{7M^2}{{\cal R}^2}\right)
iv\gamma^2 - \frac{M^2
}{{\cal R}^2}\,\frac{v\gamma^2}{{\cal R}^2
}
\frac{\chi^4}{4a^2} \sin(2\Theta)\cos{\Phi}\left[
{\rm sgn}[\cos{\Theta}]\cos{2\Theta}-i\gamma^2
\sin{2\Theta}\cos{\Phi}
\right]\right]{\rm sgn}[\cos{\Theta}]
\end{multline}
\begin{equation}\label{adap-00} 
\widetilde{\Psi}^{\bf 00}=\left(\frac{4M}{\cal R}+\frac{7M^2}{{\cal R}^2}\right)
 \frac{v^2\gamma^2}{\sqrt 3\;}
-\frac{M^2
}{\sqrt{3\;}{\cal R}^4}\,
\frac{\chi^4}{4a^2}
\left(1+(\gamma^4-1)\sin^2{2\Theta}\cos^2{2\Phi}
\right)
\end{equation}
\begin{equation} \label{adap-20}
\widetilde{\Psi}^{\bf 20}=-\,\left(\frac{4M}{\cal R}+\frac{7M^2}{{\cal R}^2}\right)
\,\frac{v^2\gamma^2}{\sqrt{6\;}}
+\frac{M^2}{{\sqrt{6\;}{\cal R}^4}}\,
\frac{\chi^4}{4a^2}
\left(\cos^2{2\Theta}+\gamma^4\sin^2{2\Theta}\cos^2{\Phi}
-2\sin^2{2\Theta}\sin^2{\Phi}\right)
\end{equation}
\begin{equation}\label{adap-21} 
\widetilde{\Psi}^{\bf 21}=
\frac{M^2}{{\cal R}^4}\,
\,
\frac{\chi^4}{4a^2}
\sin{2\Theta}\sin{\Phi}\left(\cos{2\Theta}\,{\rm sgn}[\cos{\Theta}]-i\,\gamma^2
\sin{2\Theta}\cos{\Phi}
\right)
\end{equation}
\begin{equation} \label{adap-22} 
\widetilde{\Psi}^{\bf 22}=
-\left(\frac{4M}{\cal R}+\frac{7M^2}{{\cal R}^2}\right)\frac{v^2\gamma^2}{2}
-\,\frac{M^2}{2{\cal R}^4}
\frac{\chi^4}{4a^2}\left(\cos^2{2\Theta}-\gamma^4\sin^2{2\Theta}\cos^2{\Phi}
-{\rm sgn}[\cos{\Theta}]
2i\gamma^2\cos{2\Theta}\sin{2\Theta}\cos{\Phi}\right)\,.
\end{equation}
Here, ${\cal R}$, in terms of adapted coordinates, is given by
\begin{equation}
{\cal R}^2
\equiv
\frac{\chi^4
}{4a^2
}\left[1+\gamma^2v^2\sin^2{2\Theta}\cos^2{\Phi}\right]\,.
\end{equation}

In principle, this completes the specification of the problem to be
computed in adapted coordinates. We summarize that problem here.
({\em i})~Our field equations are those of Eq.~(\ref{boxComplex}), for
$A=(nn),(n0),(00), (20)$ and Eqs.~(\ref{BoxU}), (\ref{BoxV}) for
$A=(n,\pm1),(2,\pm1),(2,\pm2)$.  In these equations, $\Box $ and
$\partial_\phi$ are given in adapted coordinates by
Eqs.~(\ref{waveq}), (\ref{FDMoutbc}).  ({\em ii})~The source terms
${\cal Q}^A$ are given, schematically, by Eq.~(\ref{RAsimple}), with
indices $\tilde{a}, \tilde{b} $ taken as adapted coordinate
labels. ({\em iii}) The outer boundary condition is given by
Eq.~(\ref{sommerfeldhelical}). ({\em iv}) The inner boundary 
conditions on the fields $\widetilde{\Psi}^A$ are given by Eqs.~(\ref{adap-nn})
-- (\ref{adap-22}).

\section{Numerical methods and results}\label{sec:numresults}

It was explained at the end of Sec.~\ref{sec:helicalscalar} that 
we work in the approximation of small orbital velocity $v$.
Of the fields $\widetilde{\Psi}^{A}$, only $\widetilde{\Psi}^{(nn)}$
requires the solution of a nonlinear problem in that small $v $ limit.
For this reason we shall emphasize, in the presentation of the
results, those for $\widetilde{\Psi}^{(nn)}$.

The nonlinear problem is that defined, for $A=(nn),
$ in
Eqs. (\ref{boxComplex}) and (\ref{RAsimple}).
More specifically, it is 
\begin{equation}\label{boxpsinn} 
	\Box\tPsi^{(nn)} = \Bigl( \tfrac{7}{8} \eta^{\kappa\nu} 
		+ \tfrac{1}{4} n^\kappa n^\nu \Bigr)\, \tPsi^{(nn)}_{,\kappa}\, \tPsi^{(nn)}_{,\nu}
+ n^\kappa n^\nu\tPsi^{(nn)}   \tPsi^{(nn)}_{,\kappa\nu}\,.
\end{equation} 
The orbital velocity $v$ is related to the mass $M
$ of either of the orbiting masses by 
\begin{equation}\label{kepler} 
v^2={M}/{4a}\,,
\end{equation}
and to linear (i.e.\,, lowest) order $\widetilde{\Psi}^{(nn)}\sim v^2$.
Here we compute $\widetilde{\Psi}^{(nn)}$ correct to second order, i.e.\,,
to order $v^4$. 

Inner boundary conditions are imposed on an approximately spherical
small surface at $\chi=\chi_{\rm min}$.  The fact that we are using
a second-order post-Minkowski approximation puts a significant
restriction on the value of $\chi_{\rm min}$.  From the definition of
$\chi$ in Eq.~(\ref{chiofXYZ}) (see also \cite{eigenspec,lineareigen}) it
follows that at a small distance $R
$
from
one of the source objects
\begin{equation}
\chi\approx\sqrt{2aR\;}\,.
\end{equation}
From this and Eq.~(\ref{kepler}) we have that the maximum field strength 
for the domain of computation, the field strength at the inner boundary, 
is 
\begin{equation}\label{MbyR} 
\left.\frac{M}{R}\right|_{\rm max}
=
\frac{8a^2
}{\chi_{\rm min}^2
}\,v^2
=
\frac{8a^4
}{\chi_{\rm min}^2
}\,\Omega^2\,.
\end{equation}
For the second-order post-Minkowksi approximation to be justified, this
measure of field strength must be significantly less than unity, so 
for a given choice of $\Omega
$ we must in principle choose the location of the inner boundary to satisfy
\begin{equation}
\chi_{\rm min}/a\gg 2\sqrt{2\;}\,a\Omega=\chi_{\rm min}/a\gg 2\sqrt{2\;}\,v\,.
\end{equation}
In the results to be presented we will vary $\Omega $ and 
$\chi_{\rm min}$ to achieve different values of $\left.M/R\right|_{\rm max}$. 
This will allow us to compare the computed errors due to the
post-Minkowski truncation with the expected errors.

An estimate of errors will be possible because we will present the
results of three different approaches to the computation of
$\widetilde{\Psi}^{(nn)}$: 

(i)~A computation using linearized general relativity, precisely the
computation done in Ref.~\cite{lineareigen}. This computation is done
by solving $\Box\widetilde{\Psi}^{(nn)}=0$ for the lowest order (
$\widetilde{\Psi}^{(nn)}=4M/{\cal R}$) boundary conditions in
Eq.~(\ref{adap-nn}), imposed at $\chi_{\rm min}$.  This should give
$\widetilde{\Psi}^{(nn)}$ correct to first order in
$\left.M/R\right|_{\rm max}$.

(ii)~A computation using the perturbative post-Minkowski approach, as
indicated in Eq.~(\ref{order2}). This approach starts with the
linearized computation of (i) above to find
$\tPsi[1]^{(nn)}$. That first order result is then used 
in Eq.~(\ref{boxpsinn}) in the 
form
\begin{equation}
\Box \tPsi[2]^{(nn)}
=\Bigl( \tfrac{7}{8} \eta^{\kappa\nu} 
		+ \tfrac{1}{4} n^\kappa n^\nu \Bigr)\, \tPsi[1]^{(nn)}_{,\kappa}\, 
\tPsi[1]^{(nn)}_{,\nu}
+ n^\kappa n^\nu\tPsi[1]^{(nn)}   \tPsi[1]^{(nn)}_{,\kappa\nu}\,.
\end{equation}
This equation is solved for 
$\tPsi[2]^{(nn)}$, using 
only the second-order part of the boundary condition in 
Eq.~(\ref{adap-nn}). (That is, the linearized part of the boundary condition,
i.e.\,, the part of the boundary condition  used for $\tPsi[1]^{(nn)}$,
is subtracted.) 
The numerical result involves the same ``eigenspectral'' method as was used
in 
Ref.~\cite{lineareigen}, with the addition of the driving terms. 
The final result is
\begin{equation}
\tPsi^{(nn)}=\tPsi[1]^{(nn)}
+\tPsi[2]^{(nn)}\,.
\end{equation}
This result is expected  to be correct to second order 
in $\left.M/R\right|_{\rm max}$.

(iii)~A computation using the exact post-Minkowski approach. Here
Eq.~(\ref{boxpsinn}) is solved as a nonlinear equation, and the full boundary
condition in 
Eq.~(\ref{adap-nn})
is used. The equation is solved by a multidimensional Newton-Raphson scheme
similar to that used for the nonlinear models in Ref.~\cite{eigenspec}. 
The result of this computation is expected to be 
$\widetilde{\Psi}^{(nn)}$
correct to second order in $\left.M/R\right|_{\rm max}$. Since this 
exact PM approach, and the PPM approach described above,
are both second order computations, we expect the difference between the 
two methods to be third order in $\left.M/R\right|_{\rm max}$.

\begin{figure}[ht] 
\includegraphics[width=.3\textwidth]{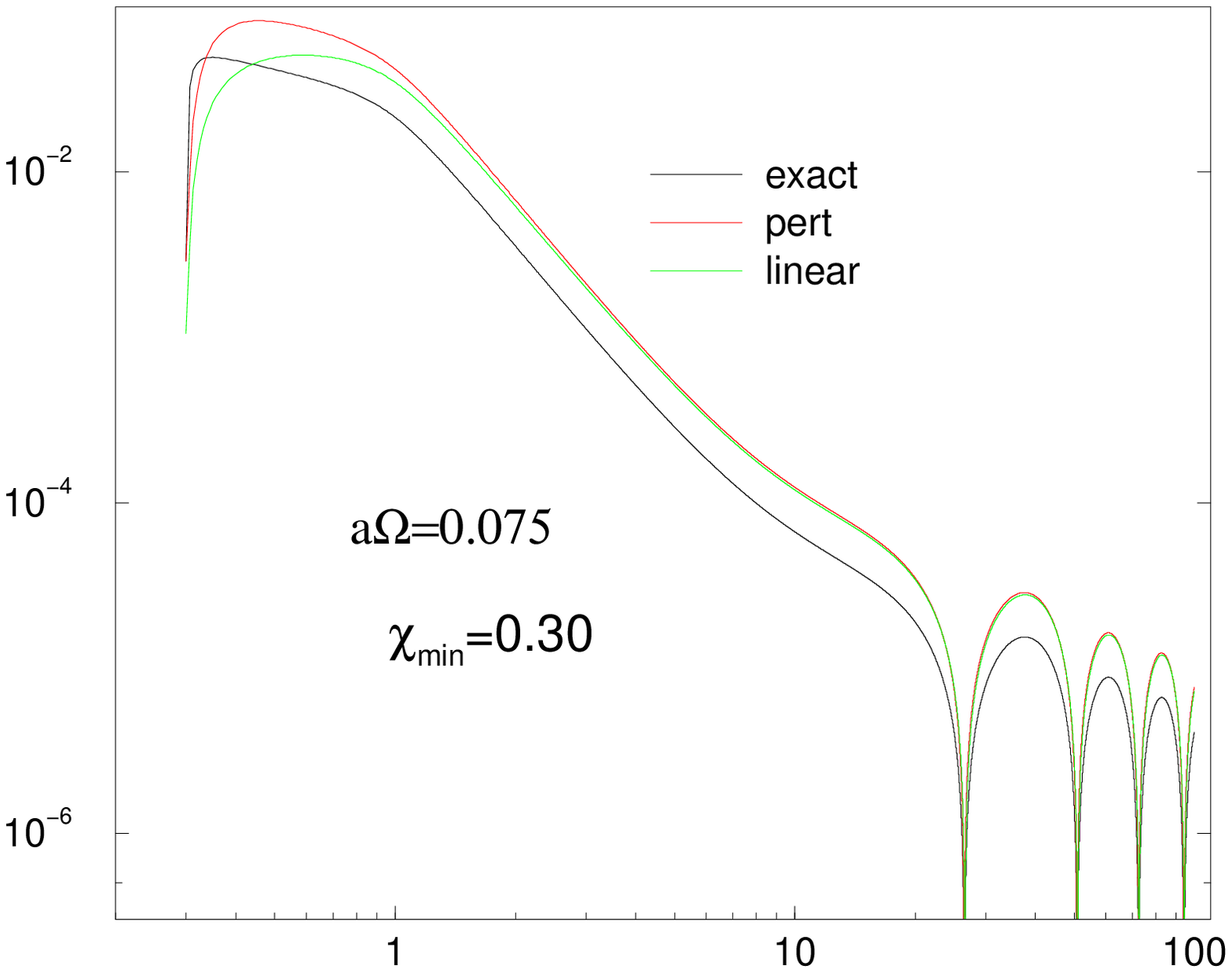}
\includegraphics[width=.3\textwidth]{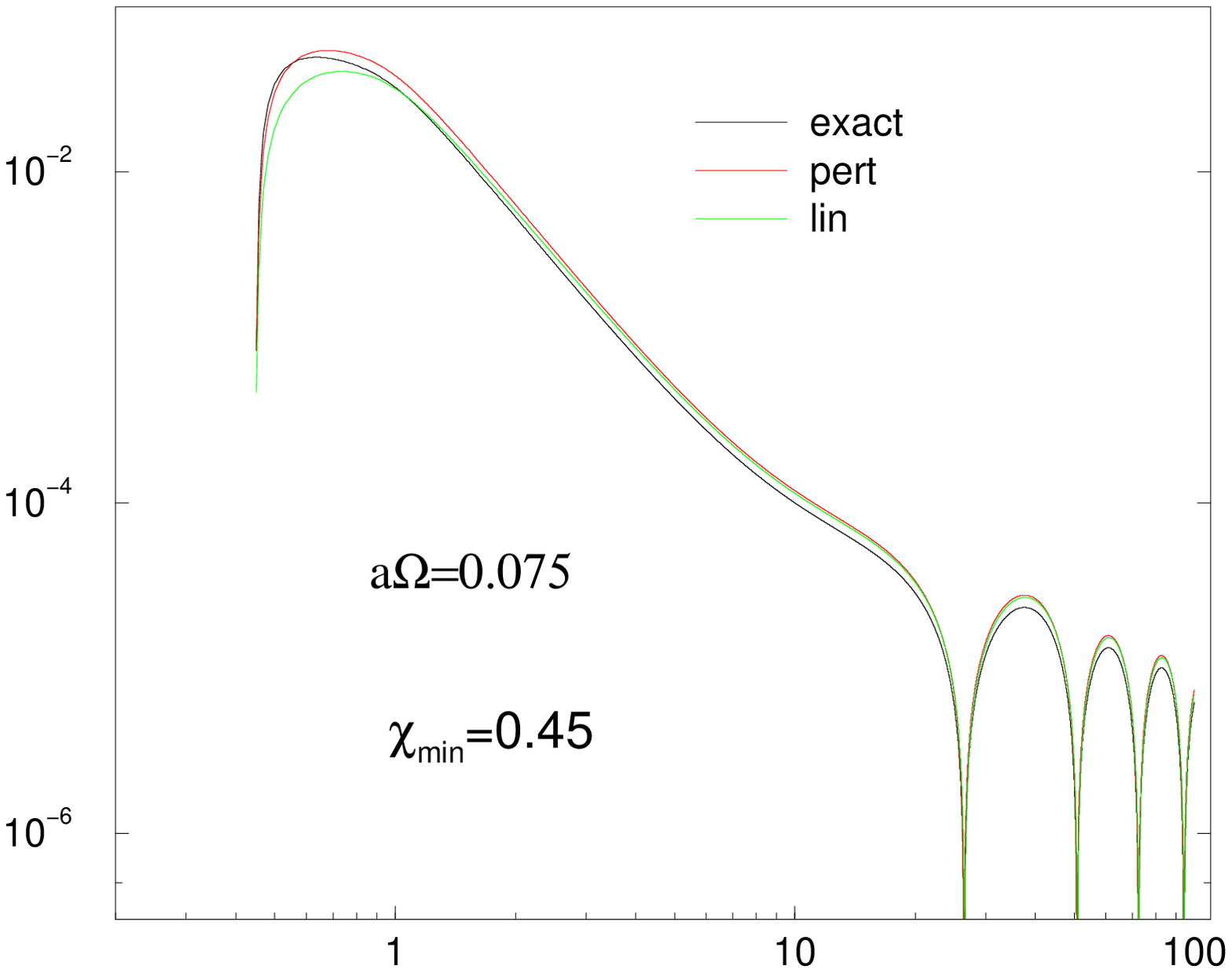}
\includegraphics[width=.3\textwidth]{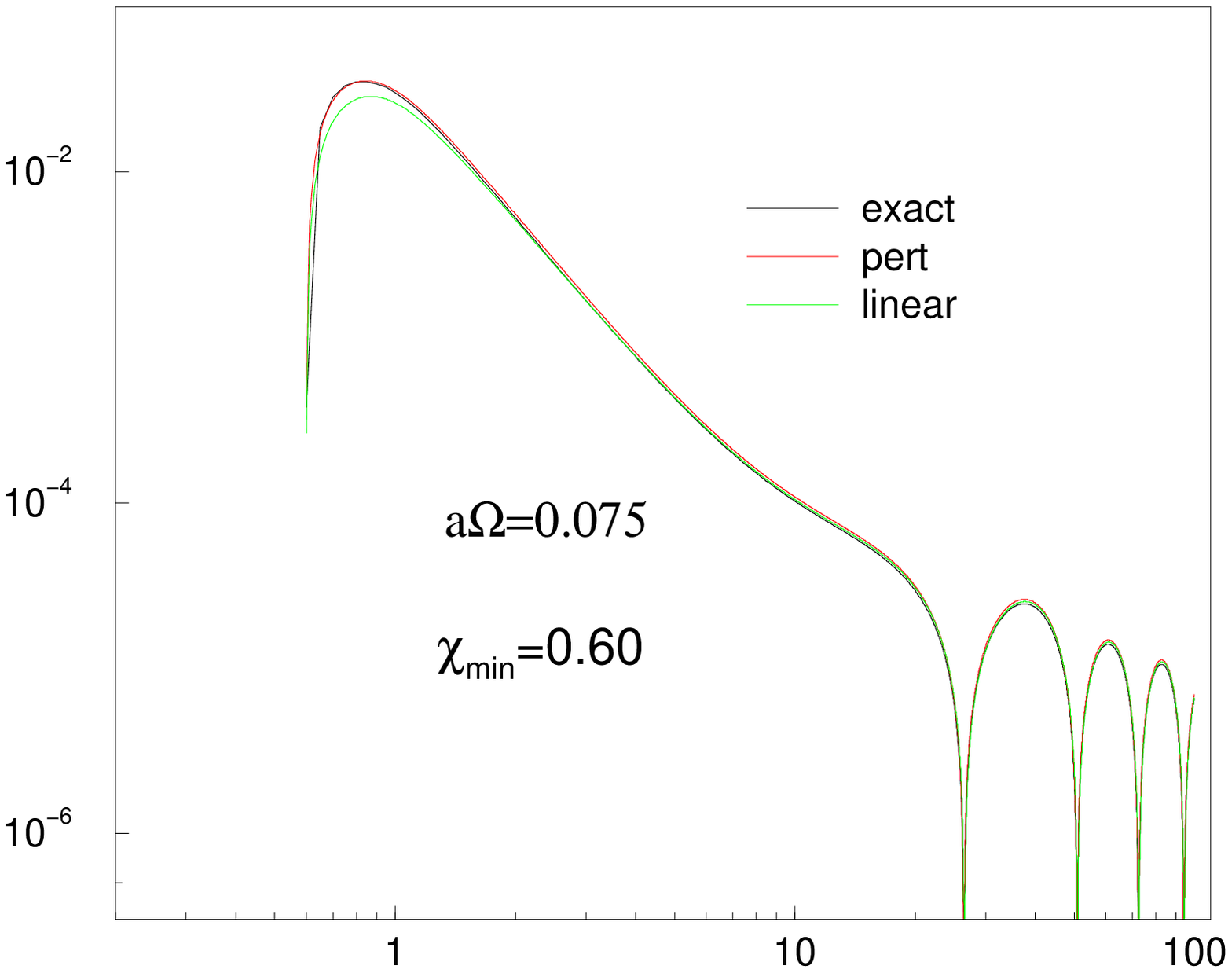}\\
\includegraphics[width=.3\textwidth]{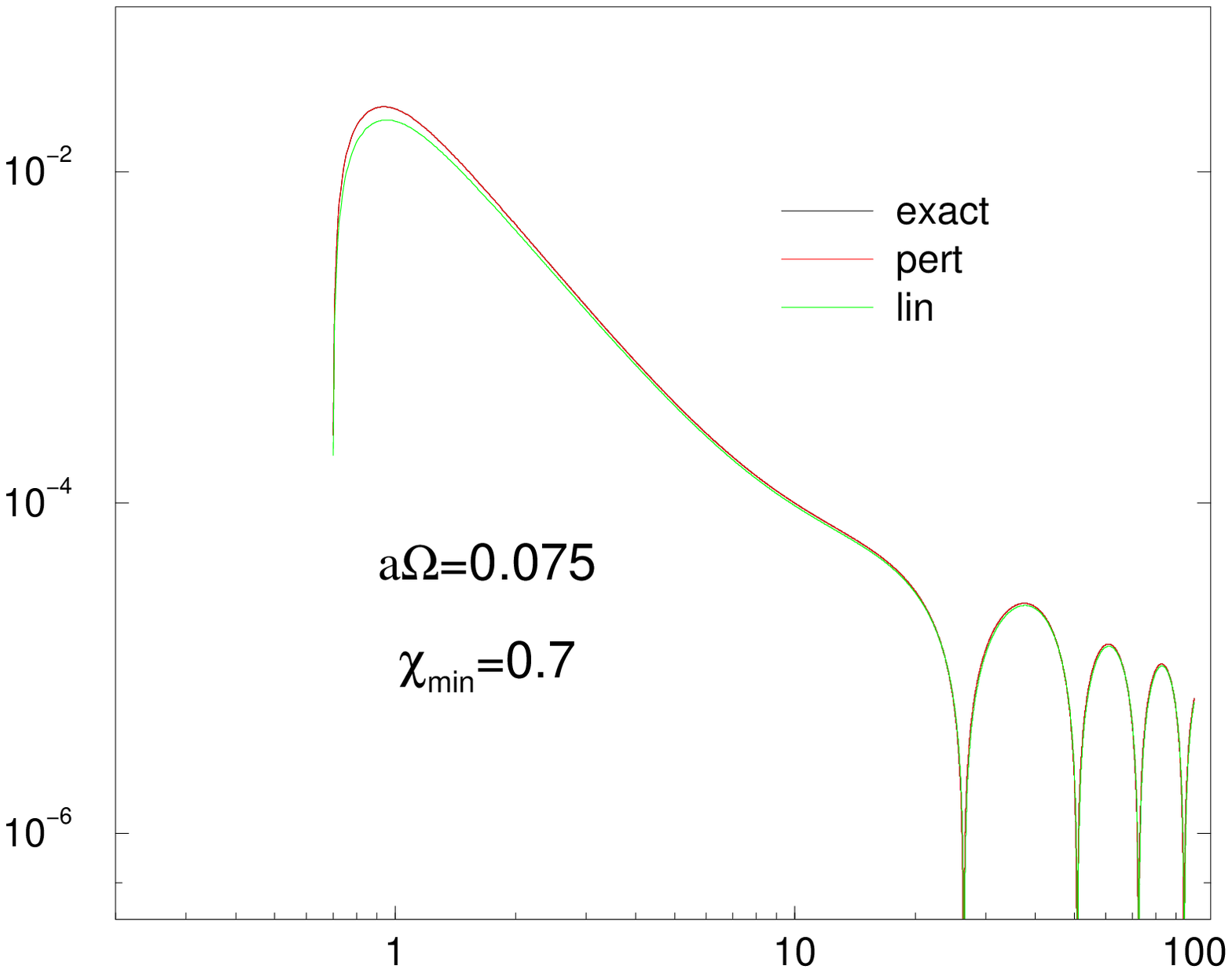}
\includegraphics[width=.3\textwidth]{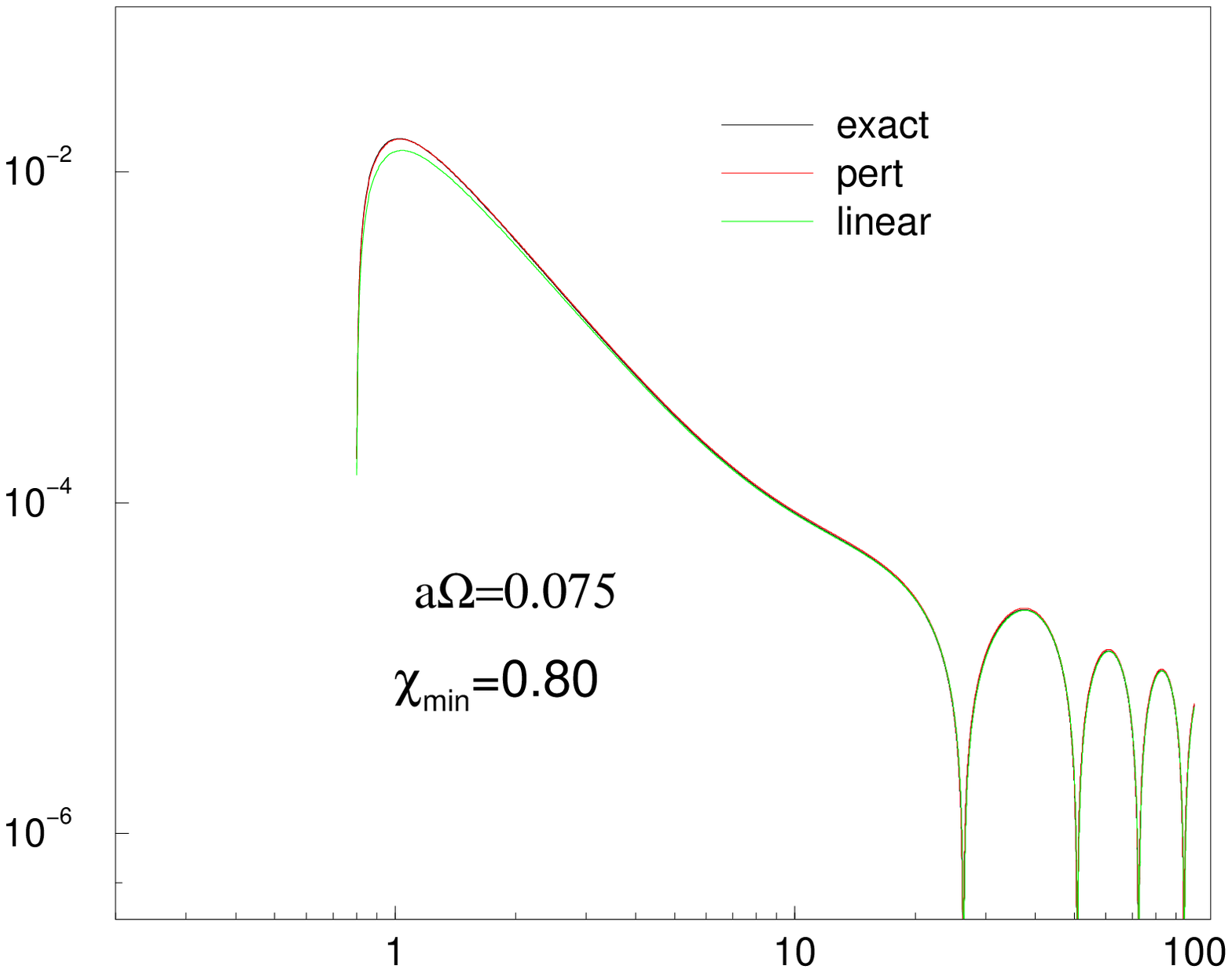}
\caption{Comparison of results for linearized, perturbative
post-Minkowski, and exact post-Minkowski computations of
$\tilde{\Psi}^{(nn)} $, as functions of $\chi/a $, in the case
$a\Omega=.075 $, for five different values of $\chi_{\rm min}$.}
\label{fig:.075x3}
\end{figure}

\begin{figure}[ht] 
\includegraphics[width=.4\textwidth]{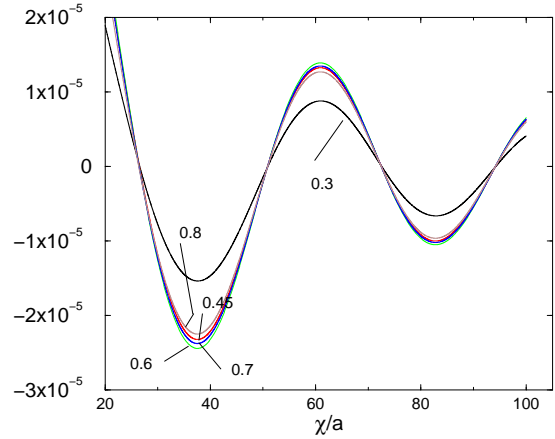}
\caption{Comparison of exact post-Minkowski computations
of $\tilde{\Psi}^{(nn)} $,
for $a\Omega=0.075
$, for five different locations of the inner boundary $\chi_{\rm min}/a$.
Each curve is labeled with the values of $\chi_{\rm min}/a$ (0.3, 0.45, 0.6, 0.7 0.8) for which 
it was computed.}
\label{fig:5way}
\end{figure}

\begin{figure}[ht] 
\includegraphics[width=.41\textwidth]{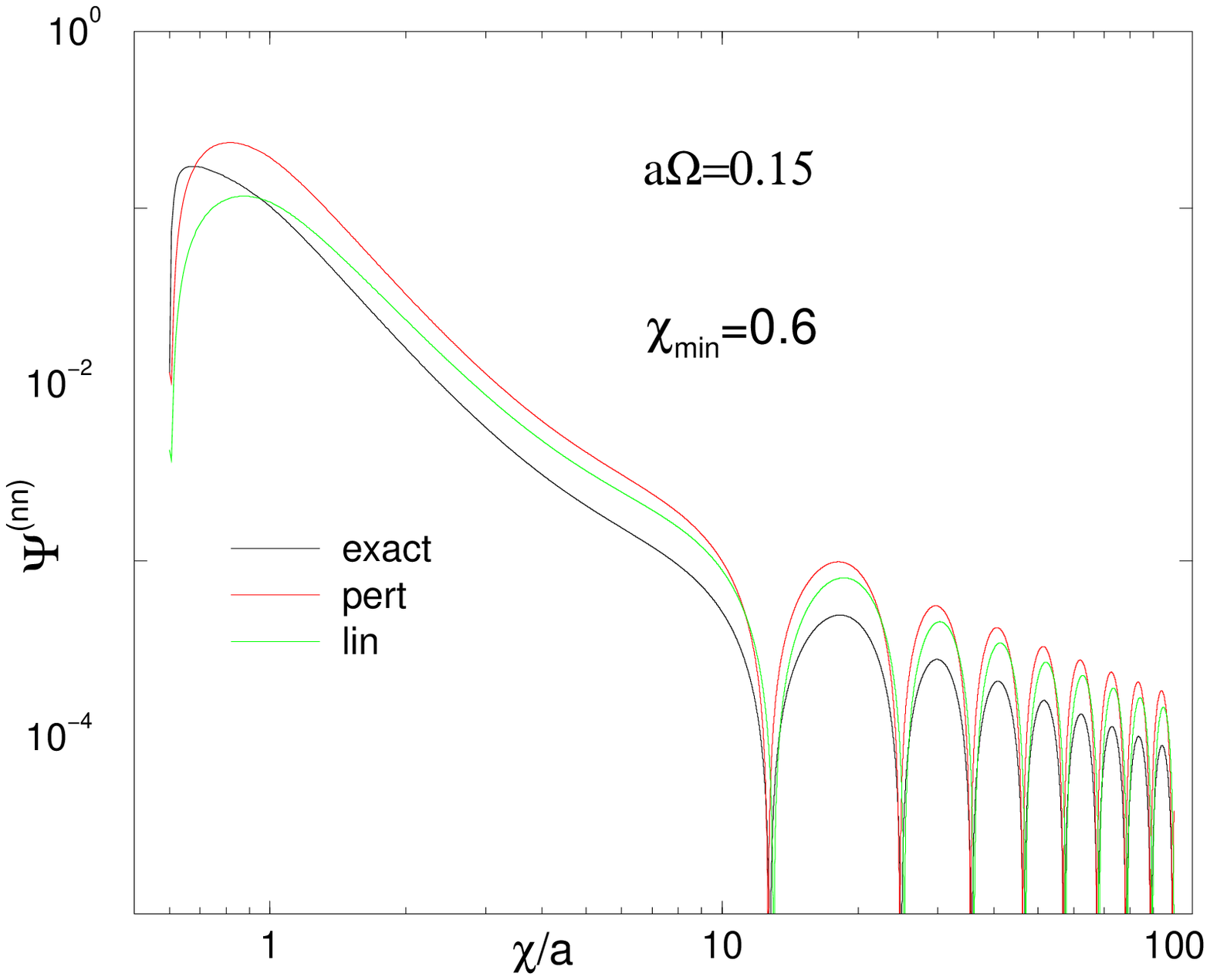}\hspace{.1in}
\includegraphics[width=.39\textwidth]{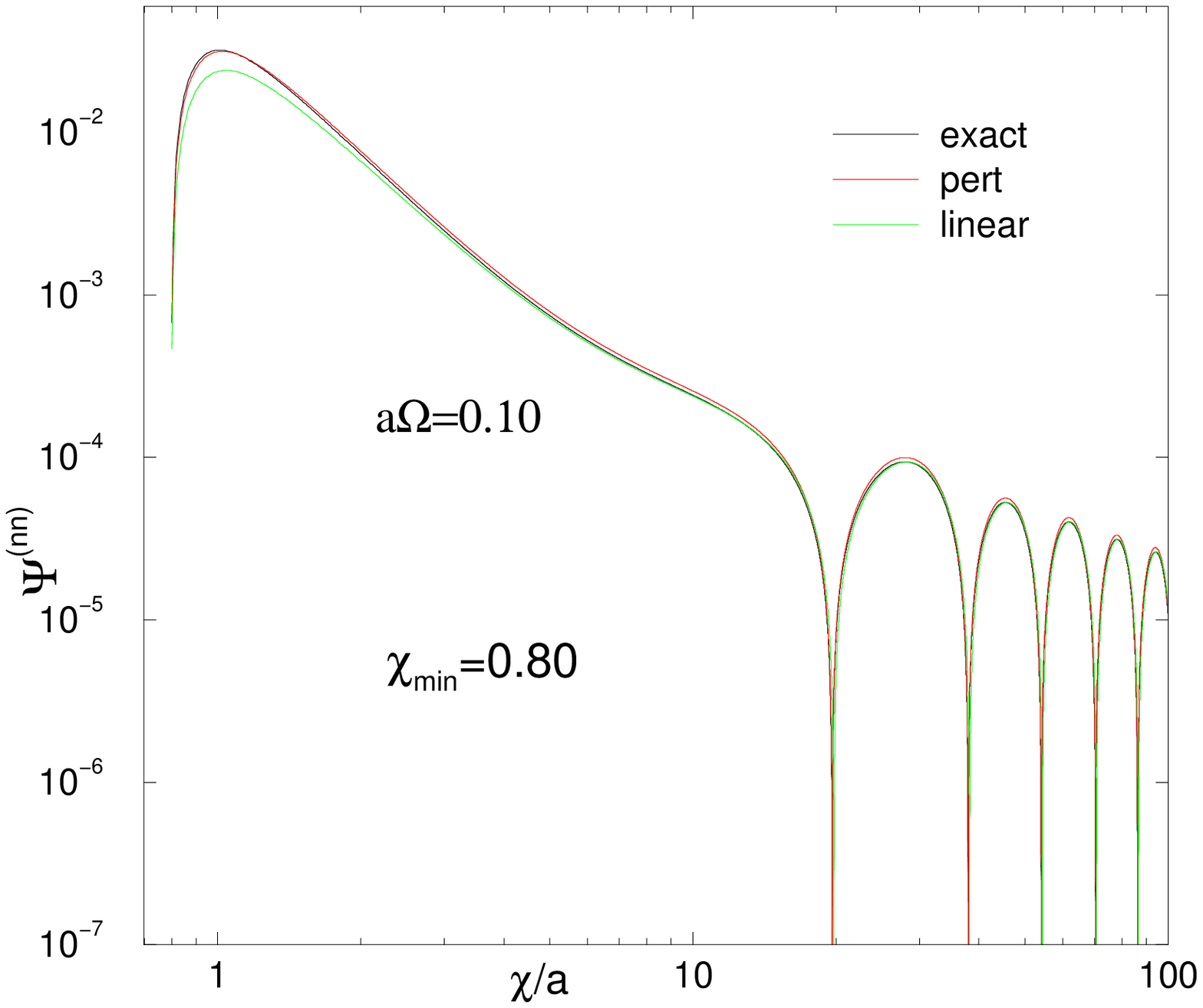}
\caption{Comparison of results for linearized, perturbative post-Minkowski,
and exact post-Minkowski computations for higher velocities.}
\label{fig:.1and.15}
\end{figure}

The results for several models are shown in Figs.~\ref{fig:.075x3} --
\ref{fig:.1and.15}.  Since the monopole moment is much larger than the
radiation fields, these figures show the computed results with the
monople moment subtracted. The sharp feature near $\chi_{\rm min}$
shows that the inherent quadrupole moment imposed by the inner
boundary conditions is immediately overwhelmed by the quadrupole
moment due to the binary configuration. This near irrelevancy of the
inherent source structure has been discussed in detail in Sec.~V, and
Appendix B, of Ref.~\cite{eigenspec}.

Each of  Figs.~\ref{fig:.075x3} --
\ref{fig:.1and.15}
corresponds to a choice of binary
velocity $v=a\Omega$, and of the inner boundary parameter $\chi_{\rm
min}$.  
All computations were done with a $16\times32\times16000
$ grid respectively in $\Theta,\Phi,\chi
$, and with six (discrete) spherical harmonics.  The Newton-Raphson 
iteration of the exact PPM computations were iterated 20 times, although
convergence was typically achieved after only three or four iterations.
Each plot shows the results of three different computations: those for
linearized theory, perturbative post-Minkowski, and exact
post-Minkowski.  Figures \ref{fig:.075x3} and \ref{fig:5way}, for
orbital velocity $v/c=0.075$, shows the effect of varying the choice
of $\chi_{\rm min}/a$ through the set of values $0.3, 0.45, 0.6, 0.7,
0.8$.  The good agreement of all three (exact, perturbative, and
linearized) computations for $\chi_{\rm min}=0.6, 0.7$, and the
excellent agreement for $\chi_{\rm min}=0.8$ is an indication that for
these choices the inner boundary is large enough that the field
strength is small at the inner boundary, and the post-Minkowski
approach is justified. For $\chi_{\rm min}=0.3$ there are significant
disagreements among the results of the three approaches. For this
case, the crude estimate in Eq.~(\ref{MbyR}) gives a value of 0.5 for
$M/R $, too large for a weak field approximation to be reliable.

Figure \ref{fig:5way} shows a detail in the wave zone, for the five 
$v/c=0.075$ exact computations. The results shows that the strong field
error for $\chi_{\rm min}/a=0.3$ significantly reduces the wave strength.
For the larger values of $\chi_{\rm min}/a$ the differences in wave strength 
are small, and have no pattern, indicating that the error is being dominated
by sources other than the strong field at the inner boundary. 
Figure \ref{fig:.1and.15} shows computational results for the larger
orbital velocities $0.1 $ and $0.15$.  In these examples we see
further evidence that higher orbital velocity, and smaller $\chi_{\rm
min}$ leads to disagreement of the three (exact, perturbative, linear)
computations, suggesting a violation of the weak-field requirement.

Table \ref{table:fieldstrength} gives a crude analysis of the
correlation of the field strength at the inner boundary and the
agreement of the three computations for a given model. The parameters
of the model are given in columns two and three with the fourth column
giving the estimate of the field strength at the inner boundary,
according to the criterion in Eq.~(\ref{MbyR}). The remaining columns
give various indicators of agreement of the linearized, PPM, and Exact
PM computations.

In the near zone the comparison is done by using differences in the
value of the first maximum of $\Psi^{(nn)}$ near the inner boundary;
this is the maximum of each of the curves of Figs.~\ref{fig:.075x3}
and \ref{fig:.1and.15}.  (The value at the inner boundary itself is
fixed by the boundary conditions.  At the inner boundary, therefore,
the results for PPM and Exact PM must be the same; the difference
between these two and the linearized computation would only reveal the
second order difference in the bounary values used.)  The columns
labeled ``Lin v. PPM'' gives the fractional difference of the computed
value of this maximum for the linearized and the PPM computation.  The
column ``PPM v. Exact'' does the same for the two second-order
post-Minkowski computations. The following two columns test the 
hypothesis that
the difference between the linearized and the PPM results are
second-order in the boundary-value field strength, and that the PPM
{\em vs.} Exact PM results are third order. If those order estimates
were accurate, the numerical values in columns seven and eight would
be expected to have little variation. Columns nine through twelve give
the same indicators as those in five through eight, but now for the
wave zone. In the wave zone the criterion for agreement is taken to be
the maximum of the first wave, e.g.\,, at $\chi\approx28 $ in the plot
for the $a\Omega=0.10 $, $\chi_{\rm min} =0.80 $ model in
Fig.~\ref{fig:.1and.15}.

\begin{table}[ht]
\begin{tabular}{|c|c|c|c||c|c|c|c|c|c|c|c|}
\hline
\multicolumn{4}{|c}{\ }&
\multicolumn{4}{|c|}{near zone}&\multicolumn{4}{c|}{wave zone}
\\ \hline
&&&&
Lin v.&
PPM v&
Lin/PPM&
Lin/Exact&
Lin v.&
PPM v&
Lin/PPM&
Lin/Exact
\\ 
model&$\Omega$&$\chi_{\rm min}$&$M/R$&
PPM&Exact&$\times(R/M)^2$
&$\times(R/M)^3$&
PPM&Exact&$\times(R/M)^2$
&$\times(R/M)^3$\\
\hline
Ia&0.075&0.3&0.5&0.61&0.67&2.4&5.3&.031&0.85&0.12&6.8\\ \hline
Ib&0.075&0.45&0.222..&0.34&0.10&6.9&9.1&.03&0.18&0.61&16.4 \\ \hline
Ic&0.075&0.6&0.125&0.24&0.006&15&2.9&.032&0.64&2.0&33 \\ \hline
Id&0.075&0.7&0.0918&.20&0.0016&24&2.1&.03&0.03&3.6&39 \\ \hline
Ie&0.075&0.8&0.0703&0.17&0.006&34&18&.029&0.020&6&58 \\ \hline\hline
II&0.100&0.8&0.125&0.31&0.019&20&9.7&.067&0.063&4&32 \\ \hline\hline
III&0.150&0.6&0.5&1.0&0.36&4&2.9&.23&1.0&9&8 \\ \hline
\end{tabular}
\caption{Agreement of the computations and field strength at the 
inner boundary. See text for details.}
\label{table:fieldstrength} 
\end{table}
The values in Table \ref{table:fieldstrength}, for the near 
zone, show weak evidence for
the expected effects of the field strength at the inner boundary
The values in the wave zone show less evidence and, along with Fig.~\ref{fig:5way},
suggest that the error in the wave zone tends not to be dominated by 
the location of $\chi_{\rm min}$.

The lack of clear evidence of field-strength effects is not
a complete
surprise.  For large values of $M/R$, the field strength for these models, we
should expect the post-Minkowski approximation to be too crude. The
disagreements on the order of 100\% are in accord with this, and an analysis
based on orders of a small parameter should fail. 
For significantly smaller  values of $M/R$ the error induced by the location 
of $\chi_{\rm min}$ are small, and other sources of error dominate.


\section{Summary and conclusions}\label{sec:conc}

In this paper we present methods and results for the post-Minkowski
level of computations within the Periodic Standing Wave approximation.
More specifically, we  present the theoretical background for
the PM computations along with two distinct ways of computationally
treating the second-order PM equations: (i)~the perturbative PM method in 
which first-order fields are initially computed and used as sources 
in the linear equations for the second order fields, and (ii)~the exact 
PM method in which the second-order PM equations are solved as if they 
constituted an exact theory.

Although the present paper deals with the second-order PM approximation,
the full equations of general relativity, in the harmonic gauge, were
given in a form easily adaptable to helical symmetry. In fact, an 
important point made in the paper is that the computational structure of 
the exact PM problem differs very little from that of 
full general relativity.

In the development of the infrastructure for for the problem, we
showed the utitility of the formalism of the helically symmetric
complex field projections $\widetilde{\Psi}^A $, introduced earlier
for linearized gravity theory\cite{lineareigen}. That formalism serves
well not only for the second-order PM work, but also for full general
relativity.

The details of the mathematical description of fields and motions
clarify the difference between the PM approximation for a binary
system and a post-Newtonian approximation for that system.  At a
characterisitic distance $R$ from one of the sources, the
gravitational field strength is of order $M/R$. Our PM approximation
demands that $M/R\ll 1$, since it was the very smallness of this term
that justifies omitting higher order terms in the PM approximation.
The PM approximation then, in a sense, is an approximation focused on
the near-source fields.  By contrast, the post-Newtonian approximation
is one in which the source velocity is small compared to $c$, or
$M/a\ll1$. In a very relativistic black hole binary, one in which the
two holes are almost in contact, there is no significant difference
between $M/R$ and $M/a$. But the PSW approximation makes no sense
unless $a/R$ is an order or magnitude or more.

Thus, in our equations, we can distinguish the approximations
associated with field strength and with small velocity, and we find it
useful to do this.  The limitation of field strength, that associated
with the PM approximation, is determined by the value of $M/R$ at the
inner boundary. The parameter for the post-Newtonian approximation, by
contrast, is simply $(a\Omega/c)^2$.  In our computations we consider
values of $M/R $ that are not exceedingly small, such as the $M/R=0.5
$ models of Table \ref{table:fieldstrength}. We may, at the same time,
keep terms that are second order in $v^2/c^2\sim M/a$, though these
terms may be orders of magnitude smaller than 0.5.  
The choice we have made in this paper is to 
keep
 terms to second order both in $M/R $ and in
$M/a$.  These distinctions will become even more important in the PSW
approximation for full general relativity.

An important aspect of the present paper is that it shows that there
are no insurmountable computational difficulties in computing the
post-Minkowski PSW fields. This more-or-less guarantees that there
will be no significant computational difficulties in the PSW problem
in full general relativity, as long as we choose the inner boundary
sufficiently large. The computational challenge for the full general
relativity problem will be in dealing with the strong fields in the case
that $\chi_{\rm min}$ is chosen small enough to give a good
representation of conditions very near a black hole. The full general
relativity problem will, of course, also entail interesting issues of
interpretation.

Lastly, the fact that we can now compute second-order PM fields
means that, in principle, our results can be used as trial initial conditions for 
numerical evolution codes. 
In practice, our results are limited to the region $\chi>\chi_{\rm min}$, and
evolution codes require data also for 
 $\chi>\chi_{\rm min}$. The obvious first attempt at a remedy to this
is to glue a pure Schwarzschild puncture into the 
 $\chi>\chi_{\rm min}$
region.

\section{Acknowledgment} 
We gratefully acknowledge the support of NSF grants 
PHY 0400588, PHY 0555644, 
PHY 0514282 and PHY 0554367,
and NASA grant ATP03-0001-0027.
We thank Kip Thorne, Lee Lindblom, Mark Scheel and the Caltech numerical 
relativity group, and John Friedman for many useful discussions and suggestions.


\appendix

\section{Details of the First Order Driving Term
}\label{app:drive}

Here we give the details of the terms on the right in Eq.~(\ref{RAsimple}).
We start with the Einstein equations truncated to second order, as given 
in Eq.~(\ref{pmtFeq}), and we write this equation as
\begin{equation}\label{barhppm} 
	\Box\, \h^{\alpha\beta} = S^{\alpha\beta\kappa\nu}_{\tau\phi\lambda\mu}
\;		\h^{\tau\phi}{}_{,\kappa}\, \h^{\lambda\mu}{}_{,\nu} +
		\h^{\rho\sigma}\, \h^{a\beta}{}_{,\rho\sigma}\,.
\end{equation}
As argued in Sec.~\ref{sec:trunc}, only the $\bar{h}^{tt}$ terms
have a nonzero first-order part, so only these need appear on the
right, and the driving term on the right in Eq.~(\ref{barhppm})
simplifies to
\begin{equation}\label{BoxhbareqS}
\Box\, \h^{\alpha\beta} = S^{\alpha\beta\kappa\nu}
\;\h^{tt}{}_{,\kappa}\, \h^{tt}{}_{,\nu} 
+		
\delta^\alpha_t\;\delta^\beta_t\;
\h^{tt}\, \h^{tt}{}_{,tt}\,,
\end{equation}
where
\begin{equation}
S^{\alpha\beta\kappa\nu}=
S^{\alpha\beta\kappa\nu}_{tttt}\,.
\end{equation}
We next substitute $\bar{h}^{\alpha\beta}=
\Psi^{A}{\bf t}^{\alpha\beta}_{\ A}
$,
from 
Eq.~(\ref{tilderepbarh}), on the left in
Eq.~(\ref{BoxhbareqS}) and use the fact that $\bar{h}^{tt}=\Psi^{(nn)}$, to get
\begin{equation}\label{BoxhbareqS2} 
	\Box\,(\Psi^{A}{\bf t}^{\alpha\beta}_{\ A}) 
=	{\bf t}^{\alpha\beta}_{\ A} \Box\Psi^{A}
= S^{\alpha\beta\kappa\nu}
\;		\Psi^{(nn)}_{,\kappa}\, \Psi^{(nn)}_{,\nu}\, +
\delta^\alpha_t\;\delta^\beta_t\;
\Psi^{(nn)}\,
\Psi^{(nn)}_{,tt}\,.		
\end{equation}

We can now use the fact that the basis tensors $
{\bf t}^{\alpha\beta}_{\ A}$
have the following orthogonality property under contraction and complex 
conjugation
\begin{equation}
\left({\bf t}^{\alpha\beta}_{\ A}\right)^*{\bf t}_{\alpha\beta\;B}=\delta_{AB}\epsilon(A)\,,
\end{equation}
where 
\begin{equation}
\epsilon(A)=\left\{
\begin{array}{ll}
-1 &\mbox{if $A=(n0), (n1), (n,-1)$}\\
+1 &\mbox{otherwise}\ .
\end{array}
\right.
\end{equation}
Using this, we contract Eq.~(\ref{BoxhbareqS2}) with $\left({\bf t}_{\alpha\beta\;A}\right)^*
$ to get
\begin{equation}
\begin{aligned}
\Box\Psi^{A} &= \Box 
\left(\widetilde{\Psi}^{A}e^{-i\mu(A)\Omega t}\right)
=e^{-i\mu(A)\Omega t}\left[
\Box\widetilde{\Psi}^{A}
-2i\mu(A)\Omega^2\partial_{\varphi}\widetilde{\Psi}^{(22)}
+\mu(A)^2
\Omega^2\widetilde{\Psi}^{A}\right] \\
&= \epsilon(A)\left({\bf t}_{\alpha\beta\;A}\right)^*\left(S^{\alpha\beta\kappa\nu}
\;		\Psi^{(nn)}_{,\kappa}\, \Psi^{(nn)}_{,\nu}\, +
\delta_{A, (nn)}
\
\delta^\alpha_t\;\delta^\beta_t\;
\Psi^{(nn)}\,
\Psi^{(nn)}_{,tt}\right)\,.	
\end{aligned}	
\end{equation}
Multiplying by $e^{i\mu(A)\Omega t}$ puts this in the form 
of Eq.~(\ref{boxComplex}), and we conclude 
\begin{equation}\label{QAexpression} 
{\cal Q}^A=e^{i\mu(A)\Omega t}
\epsilon(A)\left({\bf t}_{\alpha\beta\;A}\right)^*\left(S^{\alpha\beta\kappa\nu}
\;		\Psi^{(nn)}_{,\kappa}\, \Psi^{(nn)}_{,\nu}\, +
\delta_{A, (nn)}
\
\delta^\alpha_t\;\delta^\beta_t\;
\Psi^{(nn)}\,
\Psi^{(nn)}_{,tt}\right)\,.
\end{equation}

This result gives ${\cal Q}^A$ in terms of derivatives of the fields
$\Psi^{A}$ with respect to the $t,x,y,z$ Minkowski-like fundamental
coordinates. What is needed is the form indicated in
Eq.~(\ref{RAdef}): derivatives of the fields $\widetilde{\Psi}^{A}$
with respect to the adapted coordinates.  The change to this form
requires two transformations.  First, the derivatives with respect to
the four Minkowski-like coordinates $t,x,y,z$ must be changed to
derivatives with respect to the rotating coordinates
$\widetilde{t}\equiv t, \widetilde{x},\widetilde{y},\widetilde{z}$. In
doing this, helical symmetry is imposed by using
Eq.~(\ref{helicalreplacement}) to replace $\partial_t$.  The fact that
${\cal Q}^A$ has been constructed to be a helical scalar guarantees
that there will be no explicit time dependence in
Eq.~(\ref{QAexpression}); ${\cal Q}^A$ will be a function only of
rotating coordinates.  How this comes about is related to the fact
that $({\bf t}_{\alpha\beta\,A})^* =e^{-i\mu(A)\Omega t}
(\widetilde{\bf t}_{\alpha\beta\,A})^* $ so that
\begin{equation}\label{QAexpression2} 
{\cal Q}^A=
\epsilon(A)\left(\widetilde{\bf t}_{\alpha\beta\;A}\right)^*\left(S^{\alpha\beta\kappa\nu}
\;		\Psi^{(nn)}_{,\kappa}\, \Psi^{(nn)}_{,\nu}\, +
\delta_{A, (nn)}
\
\delta^\alpha_t\;\delta^\beta_t\;
\Psi^{(nn)}\,
\Psi^{(nn)}_{,tt}\right)\,.
\end{equation}

As an explicit example, here we  evaluate ${\cal Q}^{(n1)}$. 
The general expression in Eq.~(\ref{QAexpression2}) becomes
\begin{equation}
{\cal Q}^{(n1)}=
\left(\widetilde{\bf t}_{\alpha\beta\,(n1)}\right)^*S^{\alpha\beta\kappa\nu}
\;\widetilde{\Psi}^{(nn)}_{,\kappa}\, \widetilde{\Psi}^{(nn)}_{,\nu}\,.
\end{equation}
[Recall that $\Psi^{(nn)}=\widetilde{\Psi}^{(nn)}
$.] It is simplest to use the $t,\widetilde{x}, \widetilde{y}, \widetilde{z}
$ basis for evaluation.
From the definition in Eqs.~(\ref{tn1def}) we have that 
$\widetilde{t}_{t\tilde{x}\,(n1)}^*=1/\sqrt{2\;}$
and 
$\widetilde{t}_{t\tilde{y}\,(n1)}^*=-i/\sqrt{2\;}$
so that 
\begin{equation}
\begin{aligned}
{\cal Q}^{(n1)} &=\frac{1}{\sqrt{2\:}}\,\left[
S^{t\widetilde{x}\tilde{\kappa}\tilde{\nu}}-iS^{t\widetilde{y}\tilde{\kappa}\tilde{\nu}}
\right]
\;\widetilde{\Psi}^{(nn)}_{,\tilde{\kappa}}\, \widetilde{\Psi}^{(nn)}_{,\tilde{\nu}}\,. \\
&=-\frac{5}{4\sqrt{2\;}}\;\widetilde{\Psi}^{(nn)}_{,t}\left[
\widetilde{\Psi}^{(nn)}_{,\tilde{x}}-i\widetilde{\Psi}^{(nn)}_{,\tilde{y}}
\right]
=\frac{5\Omega}{4\sqrt{2\;}}\;\left[
\widetilde{x}\widetilde{\Psi}^{(nn)}_{,\tilde{y}}
-
\widetilde{y}\widetilde{\Psi}^{(nn)}_{,\tilde{x}}
\right]
\left[
\widetilde{\Psi}^{(nn)}_{,\tilde{x}}-i\widetilde{\Psi}^{(nn)}_{,\tilde{y}}
\right]\,,
\end{aligned}
\end{equation}
where the time derivative of $\Psi^{(nn)} $ has been replaced
following the helical prescription of Eq.~(\ref{helicalreplacement})

It is convenient next to reexpress the results in terms of the 
rotating coordinates $\widetilde{X},\widetilde{Y},\widetilde{Z}
$ in which the $\widetilde{Z}$ axis is aligned with the sources, rather than the 
rotational axis. For our example, this gives
\begin{equation}\label{Qn1XYZ} 
{\cal Q}^{(n1)}=\frac{5\Omega}{4\sqrt{2\;}}\;\left[
\widetilde{Z}\widetilde{\Psi}^{(nn)}_{,\widetilde{X}}
-\widetilde{X}\widetilde{\Psi}^{(nn)}_{,\widetilde{Z}}
\right]
\left[
\widetilde{\Psi}^{(nn)}_{,\widetilde{Z}}-i\widetilde{\Psi}^{(nn)}_{,\widetilde{X}}
\right]\,.
\end{equation}
Finally we can convert the derivatives with respect to 
$\widetilde{X},\widetilde{Y},\widetilde{Z}$ to derivatives with respect 
to the adapted coordinates $\chi,\Theta,\Phi
$ by using the following relationships (see Refs.~\cite{eigenspec},\cite{lineareigen}).
With the definitions
\begin{equation}
S_{p}\equiv\sqrt{2a^2+2\chi^2\cos{2\Theta}+2Q\;}
\quad\quad\quad
S_{m}\equiv\sqrt{-2a^2-2\chi^2\cos{2\Theta}+2Q\;}\,,
\end{equation}
the partial derivatives for transforming to adapted coordinates
take the fairly simple form
\begin{equation}
\frac{\partial\chi}{\partial X}
=\frac{
S_{m}
\cos{\Phi}
\left(a^2+Q\right)}
{2\chi^3}
\end{equation}
\begin{equation}
\frac{\partial\chi}{\partial Y}
=\frac{
S_{m}
\sin{\Phi}
\left(a^2+
Q\right)
}
{2\chi^3}
\end{equation}
\begin{equation}
\frac{\partial\chi}{\partial Z}
=\frac{
S_{p}
\left(-a^2+
Q\right)
}
{2\chi^3}
\end{equation}
\begin{equation}
\frac{\partial\Theta}{\partial X}
=\frac{
S_{p}
\cos{\Phi}
\left(-a^2+
Q\right)
}
{2\chi^4}
\end{equation}
\begin{equation}
\frac{\partial\Theta}{\partial Y}
=\frac{
S_{p}
\sin{\Phi}
\left(-a^2+
Q\right)
}
{2\chi^4}
\end{equation}
\begin{equation}
\frac{\partial\Theta}{\partial Z}
=-\,\frac{
S_{m}
\left(a^2+
Q\right)
}
{2\chi^4}
\end{equation}
\begin{equation}
\frac{\partial\Phi}{\partial X}
=-\,\frac{2\sin{\Phi}}{
S_{m}
}
\end{equation}
\begin{equation}
\frac{\partial\Phi}{\partial Y}
=\frac{2\cos{\Phi}}{
S_{m}
}
\end{equation}
\begin{equation}
\frac{\partial\Phi}{\partial Z}=0\,.
\end{equation}

The first factor in our example  in
Eq.~(\ref{Qn1XYZ}) has already appeared in Eq.~(\ref{FDMoutbc}):
\begin{equation}
\left[
\widetilde{Z}\widetilde{\Psi}^{(nn)}_{,\widetilde{X}}
-\widetilde{X}\widetilde{\Psi}^{(nn)}_{,\widetilde{Z}}
\right]=
\left(\Gamma^\Theta\frac{\partial\widetilde{\Psi}^{(nn)}}{\partial\Theta}
+\Gamma^\Phi\frac{\partial\widetilde{\Psi}^{(nn)}}{\partial\Phi}
+\Gamma^\chi\frac{\partial\widetilde{\Psi}^{(nn)}}{\partial\chi} \right)\,.
\end{equation}
The $\Gamma
$ coefficients, given in the appendix of Ref.~\cite{lineareigen},
are
\begin{equation}
\Gamma^\chi=
\left(\widetilde{Z}\frac{\partial\chi}{\partial\widetilde{X}}
-\widetilde{X}\frac{\partial\chi}{\partial\widetilde{Z}}\right)
=\frac{a^2\cos\Phi\,\sin(2\Theta)
}{\chi}
\end{equation}
\begin{equation}
\Gamma^\Theta=\left(\widetilde{Z}\frac{\partial\Theta}{\partial\widetilde{X}}
-\widetilde{X}\frac{\partial\Theta}{\partial\widetilde{Z}}\right)
=\frac{\cos\Phi\left(a^2\cos(2\Theta)+\chi^2\right)
}{\chi^2}
\end{equation}
\begin{equation}
\Gamma^\Phi=\left(\widetilde{Z}\frac{\partial\Phi}{\partial\widetilde{X}}
-\widetilde{X}\frac{\partial\Phi}{\partial\widetilde{Z}}\right)
=-\,\frac{\chi^2\sin\Phi\sin(2\Theta)}{-a^2-\chi^2\cos{2\Theta}+Q}\ .
\end{equation}


\begin{thebibliography}{10}
\bibitem{notrelated} The coordinates $X,Y,Z
$ here have no relations to the Minkowski coordinates $X,Y,Z
$ used in Refs.~\cite{paperI,eigenspec,lineareigen}.
\bibitem{noissue} That such a path is appropriate (i.e.\,, is a geodesic)
is inherent in the helical symmetry that is assumed for the solution, 
and the form of the helical Killing vector given in Eq.~(
\ref{xidef2}).
\end{thebibliography}
\end{document}